\documentclass[twocolumn]{aastex63}

\received{Dec 25, 2020}
\revised{Aug 16, 2021}
\accepted{Sep 13, 2021}
\submitjournal{ApJ}

\shorttitle{Anomalous \ion{H}{1} Line Ratios in ULIRGs}
\shortauthors{Yano et al.}

\begin{document}

\title{Anomalous Hydrogen Recombination-Line Ratios in Ultraluminous Infrared Galaxies}

\correspondingauthor{Shunsuke Baba}
\email{shunsuke.baba@nao.ac.jp}

\author{Kenichi Yano}
\affiliation{Department of Physics, Graduate School of Science, The University of Tokyo,
7-3-1 Hongo, Bunkyo-ku, Tokyo 113-0033, Japan}
\affiliation{Institute of Space and Astronautical Science, Japan Aerospace Exploration Agency,
3-1-1 Yoshinodai, Chuo-ku, Sagamihara, Kanagawa 252-5210, Japan}

\author[0000-0002-9850-6290]{Shunsuke Baba}
\altaffiliation{JSPS Fellow (PD)}
\affiliation{National Astronomical Observatory of Japan,
2-21-1 Osawa, Mitaka, Tokyo 181-8588, Japan}

\author[0000-0002-6660-9375]{Takao Nakagawa}
\affiliation{Institute of Space and Astronautical Science, Japan Aerospace Exploration Agency,
3-1-1 Yoshinodai, Chuo-ku, Sagamihara, Kanagawa 252-5210, Japan}

\author{Matthew Malkan}
\affiliation{Department of Physics and Astronomy, University of California, Los Angeles,
CA 90095-1547, USA}

\author{Naoki Isobe}
\affiliation{Institute of Space and Astronautical Science, Japan Aerospace Exploration Agency,
3-1-1 Yoshinodai, Chuo-ku, Sagamihara, Kanagawa 252-5210, Japan}

\author{Mai Shirahata}
\affiliation{Institute of Space and Astronautical Science, Japan Aerospace Exploration Agency,
3-1-1 Yoshinodai, Chuo-ku, Sagamihara, Kanagawa 252-5210, Japan}

\author{Ryosuke Doi}
\affiliation{Department of Physics, Graduate School of Science, The University of Tokyo,
7-3-1 Hongo, Bunkyo-ku, Tokyo 113-0033, Japan}
\affiliation{Institute of Space and Astronautical Science, Japan Aerospace Exploration Agency,
3-1-1 Yoshinodai, Chuo-ku, Sagamihara, Kanagawa 252-5210, Japan}

\author[0000-0002-9552-555X]{Vanshree Bhalotia}
\affiliation{Department of Physics and Astronomy, University of Hawai'i at M\={a}noa,
2505 Correa Rd, Honolulu, HI 96822, USA}

\begin{abstract}

We conducted systematic observations of the \ion{H}{1}
Br$\alpha$ (4.05~$\mu$m) and Br$\beta$ (2.63~$\mu$m) lines
in 52 nearby ($z<0.3$) ultraluminous infrared galaxies (ULIRGs) with \textit{AKARI}.
Among 33 ULIRGs wherein the lines are detected,
three galaxies show anomalous Br$\beta$/Br$\alpha$
line ratios ($\sim1.0$), which are
significantly higher than those for case~B (0.565).
Our observations also show that
ULIRGs have a tendency to exhibit higher Br$\beta$/Br$\alpha$
line ratios than those observed in Galactic \ion{H}{2} regions.
The high Br$\beta$/Br$\alpha$ line ratios cannot be explained
by a combination of dust extinction and case~B
since dust extinction reduces the ratio.
We explore possible causes for the high Br$\beta$/Br$\alpha$ line ratios
and show that
the observed ratios
can be explained by a combination of an optically thick
Br$\alpha$ line and an optically thin Br$\beta$ line.
We simulated the \ion{H}{2} regions in ULIRGs with the Cloudy code,
and our results show that the high Br$\beta$/Br$\alpha$ line ratios
can be explained by high-density conditions, wherein the Br$\alpha$ line
becomes optically thick.
To achieve a column density large enough
to make the Br$\alpha$ line optically thick within a single \ion{H}{2} region,
the gas density must be as high as $n\sim10^8$~cm$^{-3}$.
We therefore propose an ensemble of \ion{H}{2} regions,
in each of which the Br$\alpha$ line is optically thick,
to explain the high Br$\beta$/Br$\alpha$ line ratio.

\end{abstract}


\section{Introduction}
\label{sec:intro}

Ultraluminous infrared galaxies (ULIRGs) are characterized by
their enormous infrared luminosities $L_\mathrm{IR}$ (8--1000~$\mu$m),
which exceed $10^{12}L_\odot$ \citep{Sanders1988uig}.
Large infrared luminosity is produced by thermal radiation
from heated dust.
This indicates that powerful energy sources
are hidden behind dust.
Since the discovery of ULIRGs in the 1980s, it has been debated
whether the dominant energy source in a ULIRG is due to
starburst activity and/or
to an active galactic nucleus (AGN)
\citep[e.g.,][]{Sanders1988uig,Sanders1996lig}.
However, the large amount of dust harbored in a ULIRG
makes it difficult to investigate the energy source observationally,
and thus it is important to determine the amount of dust extinction
in a ULIRG in order to correct the observed quantities.

One of the most widely used indicators of dust extinction
is the ratio of hydrogen recombination lines.
These lines have been
extensively studied and are widely used as tracers of ionized gas
because hydrogen is the simplest and most abundant element in the universe
\citep[e.g.,][]{Seaton1959s,Johnson1972a,Hummer1987rli,Storey1995rli}.
The ratio of the \ion{H}{1} line fluxes from photoionized gas
can be calculated numerically
for the so-called the ``case~B'' model,
in which all the Lyman-line photons are
assumed to be absorbed by other hydrogen atoms,
and all other \ion{H}{1} lines are assumed to be optically thin.
The line ratios calculated for case~B are widely recognized to
explain the observed line ratios
in Galactic \ion{H}{2} regions and in nearby starburst galaxies
\citep[e.g.,][]{Osterbrock2006agn}.
Comparing the observed line ratios with those from case~B
enables us to determine the amount of dust extinction
in the \ion{H}{1} lines.
In optical-through-infrared wavelengths,
the dust extinction is larger at shorter wavelengths
\citep[e.g.,][]{Draine2003idg}.
Thus, if we determine the ratio of two \ion{H}{1} lines,
the line with the shorter wavelength is affected more by extinction than
the one with the longer wavelength, and hence
the observed line ratio
deviates from the case~B prediction.
Accordingly, we can evaluate dust extinction using the deviation
of the observed line ratio from the case~B value.

The \ion{H}{1} line ratio that is most widely used
to determine dust extinction
is the ratio of the optical H$\alpha$
and H$\beta$
lines \citep[e.g.,][]{Veilleux1995osl,Kim1998i1j},
i.e., the so-called Balmer decrement,
because the wavelengths of these lines are easily accessible from ground-based telescopes.
However, in objects such as ULIRGs, which are heavily dust-obscured,
the dust extinction is so high that
the optical Balmer lines trace only the outer regions of the object
and therefore underestimate the extinction.
To avoid this problem, we focus here on the infrared \ion{H}{1} lines
Br$\alpha$ 
($\mathcal{N}=5\rightarrow4$, 4.051~$\mu$m)
and Br$\beta$
($\mathcal{N}=6\rightarrow4$, 2.626~$\mu$m),
which are less affected by dust extinction
compared with the optical lines.
For instance,
the H$\beta$/H$\alpha$ line ratio is nearly halved
from the case B result by dust extinction of $A_V\sim1$~mag,
whereas the Br$\beta$/Br$\alpha$ line ratio is reduced by only $\sim4\%$
from case B by the same extinction \citep{Draine2003idg}.
In contrast,
dust extinction of $A_V>15$~mag
is expected in ULIRGs \citep[e.g.,][]{Genzel1998wpu};
the Br$\beta$/Br$\alpha$ line ratio is half
that of case B with this large dust extinction.
Thus, the Br$\beta$/Br$\alpha$ line ratio
is expected to be a good indicator of high dust extinction in ULIRGs.

To investigate the Br$\alpha$ and Br$\beta$ lines simultaneously,
we utilized near-infrared spectroscopy from the \textit{AKARI}
infrared satellite \citep{Murakami2007iam,Onaka2007ici}.
Owing to its unique 2.5--5.0~$\mu$m wavelength coverage,
which is not completely achievable with ground-based telescopes due to Earth's atmosphere,
we can determine the Br$\beta$/Br$\alpha$ line ratio without
observational bias such as aperture differences.
In this study,
we discuss the results of systematic observations
of the Br$\beta$/Br$\alpha$ line ratios in ULIRGs from \textit{AKARI} and
report the discovery of an anomaly in the \ion{H}{1} line ratio,
which we cannot explain with case B and dust extinction.
We also present the results of narrow-band imaging observations
of the H$\alpha$ line flux using the Nickel 40-inch telescope at Lick Observatory,
which we compare with the \textit{AKARI} results.
In Section~\ref{sec:obs}, we present our targets,
observations,
and methods of data reduction.
The resulting spectra and measured fluxes of
the Br$\alpha$, Br$\beta$, and H$\alpha$ lines
are presented in Section~\ref{sec:res}.
The observed Br$\beta$/Br$\alpha$ line ratios
are compared with those for case B,
and we conclude that some ULIRGs show an
anomalously high Br$\beta$/Br$\alpha$ line ratio.
This cannot be explained
by a combination of dust extinction and case~B
since dust extinction reduces the ratio.
In Section~\ref{sec:ioa}, we discuss
possible causes of the high Br$\beta$/Br$\alpha$ line ratio.
We find that the anomaly can be explained
with high-density \ion{H}{2} regions
which make the Br$\alpha$ line optically thick.
This high-density model is compared with
other observations of \ion{H}{1} lines in Section~\ref{sec:comp}.
Possible structures of high-density \ion{H}{2} region in ULIRGs
are discussed in Section~\ref{sec:str}.
We present some implications from our results in Section~\ref{sec:imp}
and summarize our study in Section~\ref{sec:sum}.
Throughout this paper,
we assume that the universe is flat, with
$\Omega_\mathrm{M}=0.27$,
$\Omega_\Lambda=0.73$,
and $H_0=70.4$~km~s$^{-1}$~Mpc$^{-1}$
\citep{Komatsu2011syw}.
We also assume the Milky Way dust model of \cite{Draine2003idg}
for the extinction curve.
In that model, dust extinction at the wavelengths of the
H$\alpha$, H$\beta$, Br$\alpha$, and Br$\beta$ lines
are taken to be $A_{\mathrm{H}\alpha}=0.776A_V$,
$A_{\mathrm{H}\beta}=1.17A_V$,
$A_{\mathrm{Br}\alpha}=3.56\times10^{-2}A_V$,
and $A_{\mathrm{Br}\beta}=8.19\times10^{-2}A_V$, respectively.

\section{Observations and Data Reduction}
\label{sec:obs}

First, we describe our \textit{AKARI} observations;
targets and methods of data reduction.
Then, we present details of our \textit{Nickel} observations.

\subsection{\textit{AKARI}}

\subsubsection{Targets}
\label{sec:tar}

Among the pointed observations from
\textit{AKARI}, we focused on those conducted
during the liquid-He-cooled holding period
\citep[2006 May 8 to 2007 August 26;][]{Murakami2007iam}
to obtain high-quality data.
Among those observations, we further focused on data obtained by
the mission program
``Evolution of ultraluminous infrared galaxies
and active galactic nuclei'' (AGNUL: P.I.,~T.~Nakagawa).
This is the same dataset
as that used by \cite{Yano2016s}.
The AGNUL program conducted systematic
near-infrared spectroscopic observations
of ULIRGs in the local universe.
During the liquid-He-cooled holding period,
50 near-infrared grism spectroscopic observations
of ULIRGs were conducted in this program.
The observation log and basic information,
such as redshifts and infrared luminosities
of the 50 AGNUL targets, are summarized
in Tables~1 and 2 of \cite{Yano2016s}, respectively.

In addition, we visually inspected all near-infrared spectra
obtained with the InfraRed Camera (IRC) during the liquid-He-cooled
holding period against the
``IRC Point Source Spectral Catalogue.''\footnote{The catalog is publicly available
at URL: \url{http://www.ir.isas.jaxa.jp/AKARI/Observation/update/20160425_preliminary_release.html}} 
We searched for possible targets
to be included in the present study
and found two galaxies (IRAS 09022$-$3615
and IRAS 10565$+$2448) in which
the Br$\alpha$ and Br$\beta$ lines
were clearly detected.
They were observed by
the mission program ``The nature of new
ULIRGs at intermediate redshift'' (NULIZ: P.I.,~H.~HoSeong).
To enlarge the sample size,
we added these two targets to our sample.
The observation log and basic information
for the two objects are summarized in
Appendix~\ref{sec:nuliz}.
Altogether, we analyzed the near-infrared
spectroscopic data for 52 objects using the observations described above.

\subsubsection{Reduction of Spectroscopic Data}

The near-infrared spectroscopic
observations we analyzed
were obtained with 
the IRC spectrograph \citep{Onaka2007ici}
on board the \textit{AKARI}
infrared satellite \citep{Murakami2007iam}.
We used a 1$\times$1~arcmin$^2$ window
to avoid source overlap.
The pixel scale of the \textit{AKARI} IRC was
$1\farcs46\times1\farcs46$.
We used the NG grism mode \citep{Onaka2007ici} to
obtain a 2.5--5.0~$\mu$m spectrum.
The NG grism has 
a dispersion of $9.7\times10^{-3}$~$\mu$m~pix$^{-1}$
and an effective spectral resolution 
of $\lambda/\delta\lambda\sim120$
at 3.6~$\mu$m for a point source.
We employed the observing mode IRC04,
in which one pointing comprised eight or nine
independent frames.
Thus, although we assigned only one pointing
for each ULIRG, we were able to eliminate the effects of cosmic-ray hits.
The total net on-source exposure time was $\sim6$~min for
each ULIRG.

We processed the data using
``IRC Spectroscopy Toolkit Version 20181203,''
the standard IDL toolkit prepared for the reduction of \textit{AKARI}
IRC spectra \citep{Ohyama2007nia,Baba2016r}.
The process was basically the same as that used by \cite{Yano2016s}
but we used the latest version of the toolkit.
Each frame was dark-subtracted, linearity-corrected,
and flat-field corrected.
Wavelength and flux calibrations were also made within the toolkit,
but once a spectrum was output, we manually adjusted the wavelength calibration
based on the location of the Br$\alpha$ line (\S\ref{sec:fob}).
We take the accuracy of the final wavelength calibration to be
smaller than 0.2~pixel or $\sim2\times10^{-3}$~$\mu$m.

The latest version of the tool kit,
which we used in the current study, 
incorporates
an error propagation algorithm that we have revised.
In older versions,
the flux error in the output spectrum contained
the sum of the three components:
(1) error determined from the standard deviation of the blank sky signal,
(2) error due to the uncertainty of the spectral response curve, and
(3) error caused by the wavelength calibration uncertainty.
Component (1) is statistical error and
should be included in the measurement of line fluxes and their ratios.
Component (2) has two factors:
(2a) uncertainty of scaling independent of wavelength, and
(2b) uncertainty of curve shape change depending on wavelength.
Of these, (2a) affects the measurement of line fluxes as a systematic error,
and is canceled out in the calculation of line-to-line ratios.
On the other hand, (2b) remains even in the line ratio estimate and should be included.
For component (3), the toolkit conservatively considers the accuracy of the wavelength origin
to be 1~pixel or $\sim0.01$~$\mu$m by default \citep{Ohyama2007nia}.
In this work, since we have tuned the origin as mentioned above,
component (3) is actually negligible.
Thus, since the purpose of this paper is the Br$\beta$/Br$\alpha$ line ratio,
components (1) and 2(b) should be included in the error budget,
while 2(a) and (3) are not.
The latest version of the toolkit has been revised
so that components (1), (2), and (3) can be output separately,
so (3) can be excluded.
However, it is difficult to disentangle sub-components (2a) and (2b).
Therefore, we have included components (1) and (2) (=(2a) and (2b))
in our analysis as a conservative estimate for the Br$\beta$/Br$\alpha$ ratio.

We estimated the spatial extension of an object
by stacking the spectrum along the dispersion direction
for each source.
The measured FWHM of the spatial profile
is typically $\sim4$--5~pixels,
which is consistent with the size of the point-spread function
of the \textit{AKARI} IRC in the spectroscopic mode \citep{Lorente2008aip}.
We adopted an aperture width
of 5~pixels ($=7\farcs3$) along the spatial direction
for spectrum extraction for each ULIRG.

\begin{figure}
\plotone{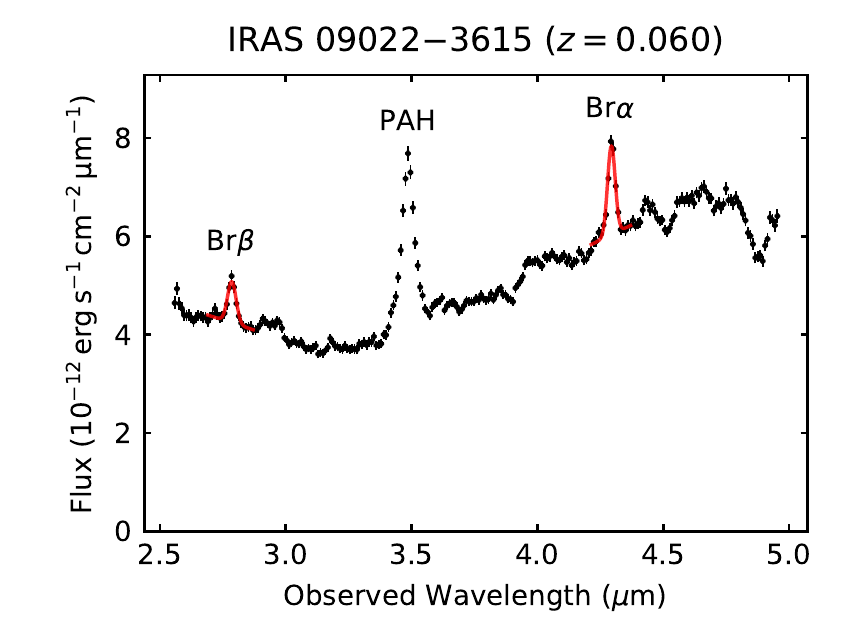}
\caption{Example of an
\textit{AKARI} IRC 2.5--5.0~$\mu$m spectrum of a ULIRG.
The best-fit Gaussian profiles for the Br$\alpha$ and Br$\beta$ lines
are plotted in red.
\label{fig:spec}}
\end{figure}

We analyzed the spectroscopic data for 52 objects,
as described in \S\ref{sec:tar}.
Since the eastern (E) and western (W) nuclei
of IRAS 17028$+$5817 are resolved by the \textit{AKARI} IRC,
the spectra of the two nuclei were extracted separately.
Thus, 53 spectra in total were obtained from the 52 observations.
Figure~\ref{fig:spec} displays a sample
2.5--5.0 $\mu$m spectrum of a ULIRG.

\subsection{Nickel 40-inch Telescope}

For two selected targets, IRAS~10494$+$4424 and Mrk~273,
which we find to show
the anomalous Br$\beta$/Br$\alpha$ ratio (\S\ref{sec:abl}),
we performed narrow-band imaging observations of the H$\alpha$
line ($\mathcal{N}=3\rightarrow2$, 6563~\AA)
with the Nickel 40-inch telescope at the Lick Observatory.
In order to compare the H$\alpha$ line flux with
the Brackett-line fluxes obtained from the \textit{AKARI} observations,
aperture matching becomes important.
Our \textit{AKARI} observations employed
slitless spectroscopy, and we used an aperture width of $\sim7''$
to extract the near-infrared spectra.
We used the narrow-band imaging observations
to obtain the H$\alpha$ line fluxes with the same aperture
as adopted for \textit{AKARI}.

We performed the Lick observations on 2014 November 26.
We used the same observational method as described in \cite{Theios2016h}.
The sky was clear, with seeing of $\sim2''$ FWHM.
We used the Nickel Direct Imaging Camera (CCD-C2),
which has $2048 \times 2048$~pixels, which we read out with $2 \times 2$ binning
to yield $1024 \times 1024$~pixels $0\farcs37$ on a side.
The observed wavelength of the H$\alpha$ line was
$7167$~\AA~at the redshift of IRAS~10494$+$4424 ($z=0.092$)
and $6811$~\AA~at that of Mrk~273 ($z=0.037$).
The filters (central wavelength/FWHM in \AA) that we used
to measure the flux of the H$\alpha$ line were
7146/80 for IRAS~10494$+$4424 and 6826/78
for Mrk~273 (on-band images).
We used the 6826 filter for IRAS~10494$+$4424
and the 7146 filter for Mrk~273
to measure the flux of the underlying continuum (off-band images).
We obtained three exposures in each of the H$\alpha$ and continuum filters,
with individual exposure times of 900~s.
We dithered the telescope between exposures to mitigate the effects
of hot or bad pixels in the detector.
We also obtained bias and twilight-sky flat-field frames for each filter.

We reduced the data using standard IRAF procedures,
including bias and flat-field corrections.
We averaged the three dithered frames in each filter
and subtracted the off-band from the on-band images to obtain
a pure H$\alpha+$[\ion{N}{2}] line image.
The narrow-band filters were photometrically calibrated
by observing standard stars and comparing them with data from the
Sloan Digital Sky Survey \citep[][]{ahn2012n}.
The fluxes through the filters were consistent with each other to within 5\%.

\section{Results}
\label{sec:res}

\subsection{Fluxes of Brackett Lines}
\label{sec:fob}

We obtained 2.5--5.0~$\mu$m near-infrared spectra
for 53 objects with \textit{AKARI}, as discussed in the previous section.
For each spectrum, the Br$\alpha$ line, at a rest-frame wavelength of
$\lambda_{\rm{rest}}=4.05$~$\mu$m,
and the Br$\beta$ line, at $\lambda_{\rm{rest}}=2.63$~$\mu$m,
were fitted separately with a linear continuum and a Gaussian profile.

First, we fitted the Br$\alpha$ line with four free parameters:
the offset and slope of the linear continuum,
the normalization of the Gaussian profile, and the central wavelength.
The line width was fixed
at the spatial width at wavelengths near Br$\alpha$
(FWHM $\sim4$--5~pixels)
since \textit{AKARI} IRC employs slitless spectroscopy
and the instrumental line spread function is determined by the spatial point spread function.
We assumed that
the spectral resolution was determined by the size of each object
because the observations employ slitless spectroscopy
and the intrinsic line widths are narrower than
the $\Delta v$ resolution of $\sim3000$~km~s$^{-1}$.
The range of wavelengths
used to fit continuum emission
was typically $\pm0.08$~$\mu$m around the central wavelength.
The central wavelengths of the Br$\alpha$ lines
exhibited small discrepancies from
those expected from the redshifts.
The discrepancy was larger than the fitting error, typically $\sim10^{-3}$~$\mu$m,
but was within the wavelength calibration error of
$\sim10^{-2}$~$\mu$m \citep{Ohyama2007nia}.
Therefore, we shifted the wavelengths for the entire spectrum
so that the best-fit central wavelength of the Br$\alpha$ line
matched the redshift.
Next, after fitting the Br$\alpha$ line, we fitted
the Br$\beta$ line, fixing the central wavelength
as expected from the redshift and
fixing the line width at the spatial width near the wavelength of Br$\beta$
as in the case of Br$\alpha$;
i.e., the free parameters are the offset and slope of the local continuum
and the normalization of the Gaussian.
We then determined the fluxes of the Br$\alpha$ and Br$\beta$ lines
by integrating the best-fit Gaussian profiles.
We show a typical result from this Gaussian fitting procedure in Figure~\ref{fig:spec}.

Among the 53 objects,
we were able to determine the fluxes
of the Br$\alpha$ and Br$\beta$ lines
for 47 galaxies.
Four of the sources (IRAS 00183$-$7111,
IRAS 04313$-$1649, IRAS 10091$+$4704, and IRAS 23498$+$2423)
have redshifts higher than 0.2;
therefore, the Br$\alpha$ line does not fall
within the 2.5--5.0~$\mu$m wavelength range.
In addition, two sources
(IRAS~21477$+$0502 and IRAS~23129$+$2548)
were found to
suffer from spectral overlapping with other objects.
In UGC~5101,
we found the continuum slope to be changing
in the vicinity of the Br$\beta$ line;
thus, we adopted a second-order polynomial
for the shape of the continuum used to fit the Br$\beta$ line
for this source.

We detected the Br$\alpha$ or Br$\beta$ lines at 
the 3$\sigma$ level in 33 objects.
For undetected lines, we derived 3$\sigma$ upper-limit fluxes.
The measured Br$\alpha$ and Br$\beta$ line fluxes
($F_{\mathrm{Br}\alpha}$ and $F_{\mathrm{Br}\beta}$) are summarized in
Table~\ref{tab:brflux}, along with the 1$\sigma$ statistical errors;
the 1$\sigma$ systematic errors were estimated to be $\sim10\%$
of the flux.

\startlongtable
\begin{deluxetable*}{ccccc}
\tablecaption{Fluxes of Brackett Lines \label{tab:brflux}}
\tablewidth{0pt}
\tablehead{
\colhead{Object Name}&\colhead{$F_{\mathrm{Br}\alpha}$\tablenotemark{a}}&\colhead{$F_{\mathrm{Br}\beta}$}&\colhead{$F_{\mathrm{Br}\beta}/F_{\mathrm{Br}\alpha}$}&\colhead{$A_V$\tablenotemark{b}}\\
\colhead{}&\multicolumn{2}{c}{($10^{-15}$ erg s$^{-1}$ cm$^{-2}$)}&\colhead{}&\colhead{(mag)}
}
\startdata
IRAS 00456$-$2904                  & 6.80$\pm$0.69  & 2.73$\pm$0.83 & 0.40$\pm$0.13   & $8.0\pm7.5$   \\
IRAS 00482$-$2721                  & 4.37$\pm$0.43  & $<$2.72       & $<$0.62         & $>-2.3$       \\
IRAS 01199$-$2307                  & 2.90$\pm$0.80  & $<$2.20       & $<$0.76         & $>-6.9$       \\
IRAS 01298$-$0744                  & $<$1.86        & $<$4.02       & \nodata         & \nodata       \\
IRAS 01355$-$1814                  & $<$2.56        & $<$2.40       & \nodata         & \nodata       \\
IRAS 01494$-$1845                  & 4.02$\pm$0.80  & 5.5$\pm$1.0   & 1.37$\pm$0.37   & $-20.7\pm6.4$ \\
IRAS 01569$-$2939                  & 4.4$\pm$1.4    & $<$4.97       & $<$1.12         & $>-16$        \\
IRAS 02480$-$3745                  & $<$4.15        & $<$2.48       & \nodata         & \nodata       \\
IRAS 03209$-$0806                  & 4.39$\pm$0.49  & 2.75$\pm$0.73 & 0.63$\pm$0.18   & $-2.4\pm6.8$  \\
IRAS 03521$+$0028                  & $<$3.63        & $<$4.38       & \nodata         & \nodata       \\
IRAS 04074$-$2801                  & $<$3.16        & 5.08$\pm$0.93 & $>$1.61         & $<-24$        \\
IRAS 05020$-$2941                  & 4.7$\pm$1.0    & $<$3.32       & $<$0.70         & $>-5.0$       \\
IRAS 05189$-$2524                  & $<$16.8        & $<$17.1       & \nodata         & \nodata       \\
IRAS 06035$-$7102                  & 8.1$\pm$1.0    & 3.80$\pm$0.93 & 0.47$\pm$0.13   & $4.2\pm6.5$   \\
IRAS 08572$+$3915                  & 25.3$\pm$2.5   & $<$5.94       & $<$0.23         & $>21$         \\
IRAS 08591$+$5248                  & 2.77$\pm$0.84  & 6.15$\pm$0.81 & 2.22$\pm$0.73   & $-32.0\pm7.7$ \\
IRAS 09022$-$3615                  & 68.8$\pm$3.3   & 35.8$\pm$2.2  & 0.520$\pm$0.041 & $1.9\pm1.8$   \\
UGC 5101                           & 18.4$\pm$2.9   & $<$40.8       & $<$2.21         & $>-32$        \\
IRAS 09463$+$8141                  & 1.86$\pm$0.51  & $<$2.54       & $<$1.36         & $>-21$        \\
IRAS 09539$+$0857                  & $<$3.36        & $<$2.13       & \nodata         & \nodata       \\
IRAS 10035$+$2740                  & $<$1.90        & $<$3.09       & \nodata         & \nodata       \\
IRAS 10494$+$4424\tablenotemark{c} & 10.92$\pm$0.83 & 10.5$\pm$1.1  & 0.96$\pm$0.12   & $-12.4\pm3.0$ \\
IRAS 10565$+$2448\tablenotemark{c} & 35.2$\pm$2.1   & 31.1$\pm$2.4  & 0.883$\pm$0.085 & $-10.5\pm2.3$ \\
IRAS 10594$+$3818                  & 8.4$\pm$1.4    & 5.51$\pm$0.93 & 0.66$\pm$0.16   & $-3.5\pm5.5$  \\
IRAS 11028$+$3130                  & $<$3.46        & $<$3.36       & \nodata         & \nodata       \\
IRAS 11180$+$1623                  & $<$4.44        & $<$2.04       & \nodata         & \nodata       \\
IRAS 11387$+$4116                  & 4.86$\pm$0.89  & $<$2.33       & $<$0.48         & $>3.9$        \\
IRAS 12447$+$3721                  & 5.63$\pm$0.80  & 3.54$\pm$0.99 & 0.63$\pm$0.20   & $-2.5\pm7.4$  \\
Mrk 231                            & $<$73.1        & $<$71.2       & \nodata         & \nodata       \\
Mrk 273\tablenotemark{c}           & 49.4$\pm$1.1   & 53.7$\pm$2.3  & 1.086$\pm$0.053 & $-15.3\pm1.1$ \\
IRAS 13469$+$5833                  & $<$3.32        & $<$4.66       & \nodata         & \nodata       \\
IRAS 13539$+$2920                  & 14.0$\pm$1.2   & 5.5$\pm$1.7   & 0.39$\pm$0.12   & $8.5\pm7.4$   \\
IRAS 14121$-$0126                  & 4.1$\pm$1.2    & $<$4.59       & $<$1.12         & $>-16$        \\
IRAS 14202$+$2615                  & 8.03$\pm$0.91  & 7.18$\pm$0.68 & 0.89$\pm$0.13   & $-10.8\pm3.5$ \\
IRAS 14394$+$5332                  & 9.5$\pm$1.1    & 8.62$\pm$0.54 & 0.91$\pm$0.12   & $-11.2\pm3.0$ \\
IRAS 15043$+$5754                  & 3.31$\pm$0.87  & 3.6$\pm$1.0   & 1.10$\pm$0.42   & $-15.6\pm8.9$ \\
IRAS 16333$+$4630                  & 4.03$\pm$0.99  & $<$3.19       & $<$0.79         & $>-7.9$       \\
IRAS 16468$+$5200                  & $<$3.16        & $<$3.24       & \nodata         & \nodata       \\
IRAS 16487$+$5447                  & 10.19$\pm$0.78 & 4.2$\pm$1.2   & 0.41$\pm$0.12   & $7.5\pm7.1$   \\
IRAS 17028$+$5817 E                & 3.80$\pm$0.69  & $<$4.88       & $<$1.28         & $>-19$        \\
IRAS 17028$+$5817 W                & 6.14$\pm$0.91  & 5.5$\pm$1.1   & 0.90$\pm$0.23   & $-11.0\pm6.0$ \\
IRAS 17044$+$6720                  & 6.6$\pm$1.1    & $<$3.95       & $<$0.60         & $>-1.3$       \\
IRAS 17068$+$4027                  & 6.53$\pm$0.90  & 3.40$\pm$0.93 & 0.52$\pm$0.16   & $1.9\pm7.2$   \\
IRAS 17179$+$5444                  & $<$3.45        & $<$4.59       & \nodata         & \nodata       \\
IRAS 19254$-$7245                  & 20.1$\pm$1.6   & 13.8$\pm$2.0  & 0.69$\pm$0.11   & $-4.6\pm3.8$  \\
IRAS 22088$-$1831                  & $<$2.58        & $<$2.35       & \nodata         & \nodata       \\
IRAS 23128$-$5919                  & 67.0$\pm$3.3   & 41.2$\pm$2.7  & 0.614$\pm$0.050 & $-2.0\pm1.9$  \\
\enddata
\tablenotetext{a}{Observed flux of the Br$\alpha$ line.
The values differ from those reported in \cite{Yano2016s}
because we used a different version of the toolkit for the data reduction
and revised the error estimation as reported in this paper.}
\tablenotetext{b}{Visual extinction derived from the observed Br$\beta$/Br$\alpha$ line ratio
assuming case~B.}
\tablenotetext{c}{Galaxies showing clear anomalies in the Br$\beta$/Br$\alpha$ line ratio.}
\end{deluxetable*}

The widths of the Br$\alpha$ and Br$\beta$ lines
are consistent with the limit of spectral resolution
($\Delta v\sim3000$~km~s$^{-1}$)
for all targets within a fitting uncertainty
of $\le100$~km~s$^{-1}$.
This indicates that
none of the objects shows a broad component
of the Br lines
with a FWHM
broader than $\sim1000$~km~s$^{-1}$.
If the broad-line region of an AGN
had contributed to the line fluxes,
the hydrogen lines would have had a FWHM
of a few thousand km~s$^{-1}$
\citep{Osterbrock2006agn}.
Thus, we conclude that the Br$\alpha$ and Br$\beta$ lines
do not originate from the broad-line regions.
In the AGNUL sample,
\cite{Yano2016s} reported a good correlation between the flux of the Br$\alpha$ line
and the 3.3~$\mu$m polycyclic aromatic hydrocarbon (PAH) emission
and suggested that any contribution from an AGN
was not dominant for the Br$\alpha$ line.
Combining these results, we conclude
that the Brackett lines originate from starburst activities in all galaxies in the sample.

\subsection{Flux of H$\alpha$ Line}

We obtained pure H$\alpha+$[\ion{N}{2}] line images
for IRAS~10494$+$4424 and Mrk~273 from the Nickel observations.
Using the IRAF phot module,
we performed circular-aperture photometry ($7''$ in diameter)
for the images and obtained the H$\alpha+$[\ion{N}{2}] line fluxes.
Assuming [\ion{N}{2}]/H$\alpha$ line ratios
to be 0.60 for IRAS~10494$+$4424 and
1.01 for Mrk~273 \citep{Kim1998i1j},
we corrected the line ratios for the [\ion{N}{2}] emission.
The resulting H$\alpha$ line fluxes
are summarized in Table~\ref{tab:haflux}.
We discuss the H$\alpha$ line flux in comparison
with the Br$\alpha$ line flux in \S\ref{sec:ioa}.

\begin{deluxetable}{cc}
\tablecaption{Observed H$\alpha$ Line Flux \label{tab:haflux}}
\tablewidth{0pt}
\tablehead{
\colhead{Object}&\colhead{$F_{\mathrm{H}\alpha}$}\\
\colhead{}&\colhead{($10^{-14}$~erg~s$^{-1}$~cm$^{-2}$)}
}
\startdata
IRAS~10494$+$4424&$1.46\pm0.15$\\
Mrk~273&$33.1\pm3.3$
\enddata
\end{deluxetable}

\subsection{Anomalous Br$\beta$/Br$\alpha$ Line Ratios}
\label{sec:abl}

Owing to the unique 2.5--5.0~$\mu$m wavelength coverage
of \textit{AKARI}, the Br$\alpha$ and Br$\beta$ lines are
observed simultaneously within a single spectrum.
This allows us to determine accurately the Br$\beta$/Br$\alpha$ line ratio
without introducing observational uncertainties such as
aperture corrections.
The estimated Br$\beta$/Br$\alpha$ line
ratios are summarized in Table~\ref{tab:brflux}.
On the basis of the Br$\beta$/Br$\alpha$ line ratio,
we were able to determine the visual extinction ($A_V$)
in the same way as for the usual Balmer-decrement method.
We assumed the intrinsic line-flux ratio for Br$\beta$/Br$\alpha$
to be 0.565 \citep[][case B with $T=10000$~K and low-density limit]{Osterbrock2006agn}.
If dust extinction affects the line fluxes, the Br$\beta$/Br$\alpha$ line ratio
decreases because the Br$\beta$ line has a shorter
wavelength and is attenuated more compared with the Br$\alpha$ line.
Thus, we expect to observe Br$\beta$/Br$\alpha$ line ratios
lower than 0.565.
In Table~\ref{tab:brflux},
we also tabulate the values of $A_V$ inferred from
the Br$\beta$/Br$\alpha$ line ratio.

\begin{figure*}
\plotone{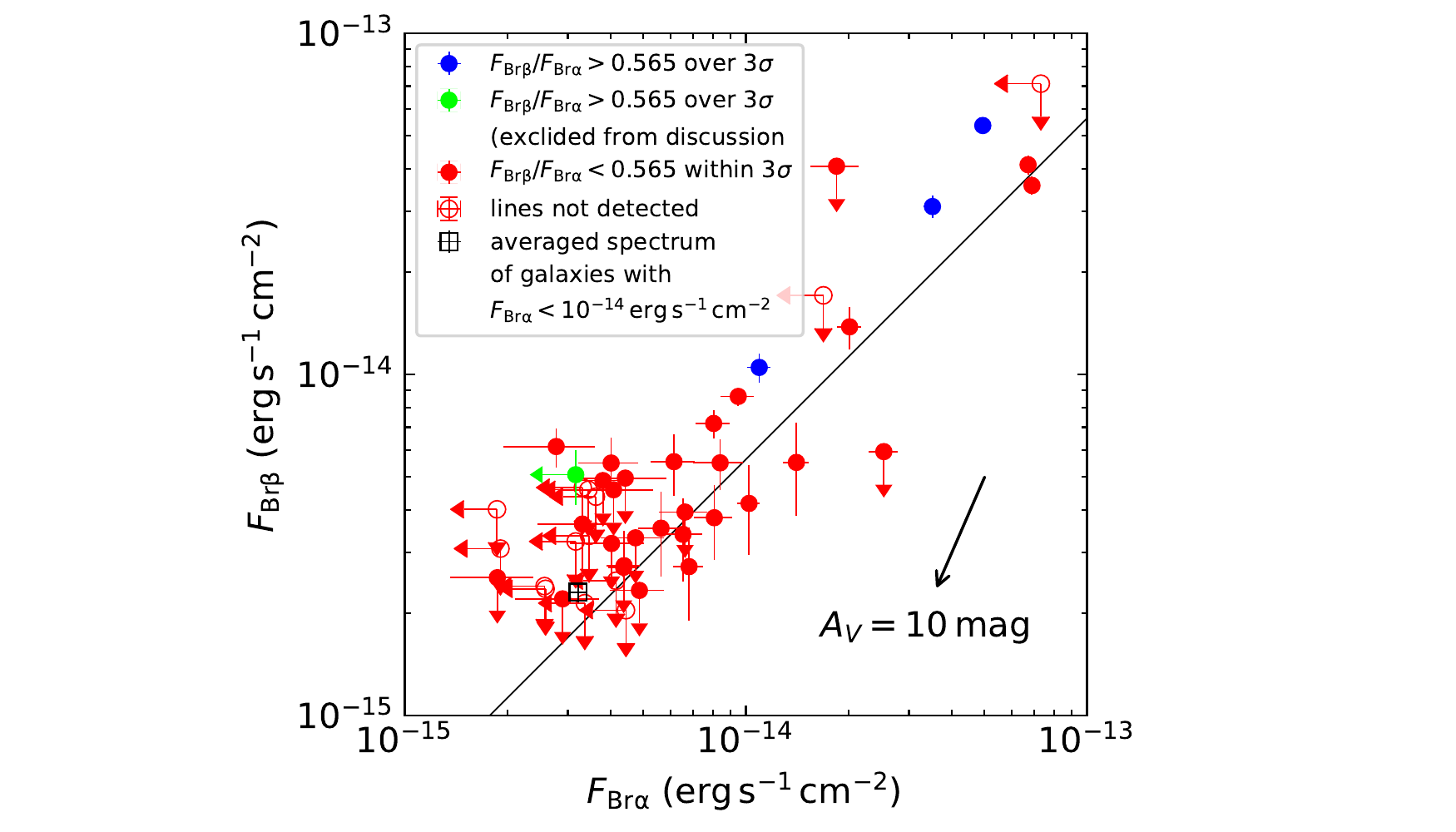}
\caption{The Br$\alpha$ line flux ($F_{\mathrm{Br}\alpha}$)
versus the Br$\beta$ line flux ($F_{\mathrm{Br}\beta}$).
The solid line shows the theoretical line ratio for case B conditions:
$F_{\mathrm{Br}\beta}/F_{\mathrm{Br}\alpha}=0.565$.
The extinction vector for $A_V=10$ mag is shown as a black arrow.
The blue filled circles show galaxies with anomalous Br$\beta$/Br$\alpha$ line ratios
(more than 3$\sigma$ higher than 0.565),
while red filled circles represent those with the normal case~B ratio.
The green filled circles show galaxies with high Br$\beta$/Br$\alpha$ line ratios
but they have been excluded from the discussion because of large uncertainties in
determining the continuum underlying the lines.
The red open circles represent galaxies where neither the Br$\alpha$ nor Br$\beta$ lines
were detected.
The black open square shows the averaged spectrum of 35 galaxies with
$F_{\mathrm{Br}\alpha}<10^{-14}$~erg~s$^{-1}$~cm$^{-2}$.
\label{fig:brflux}}
\end{figure*}

The comparison of $F_{\mathrm{Br}\alpha}$ and $F_{\mathrm{Br}\beta}$
is shown in Figure~\ref{fig:brflux}.
Galaxies located below the case B line in Figure~\ref{fig:brflux}
have a Br$\beta$/Br$\alpha$ line ratio
lower than 0.565.
The flux ratios for these galaxies are consistent with case B theory
and dust extinction,
and the $A_V$ magnitude was found to be positive for these objects.
However, for four galaxies, we obtained an anomalous
Br$\beta$/Br$\alpha$ line ratio,
which is more than 3$\sigma$ higher than 0.565.
These galaxies are located above the case B line in Figure~\ref{fig:brflux};
i.e., the Br$\beta$ line is enhanced relative to the Br$\alpha$ line.
This is opposite to the effect of dust extinction.

\begin{figure}
\gridline{\fig{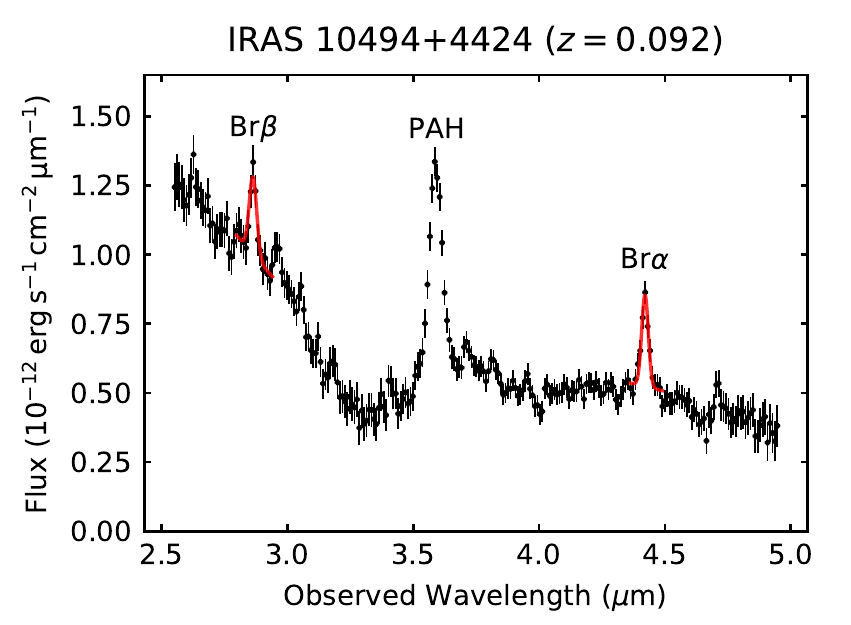}{\columnwidth}{}}
\gridline{\fig{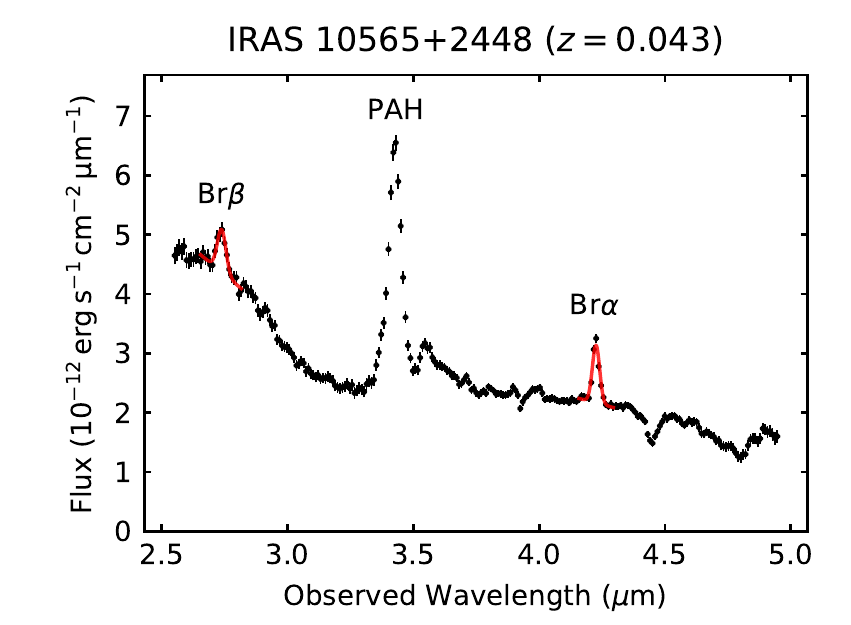}{\columnwidth}{}}
\gridline{\fig{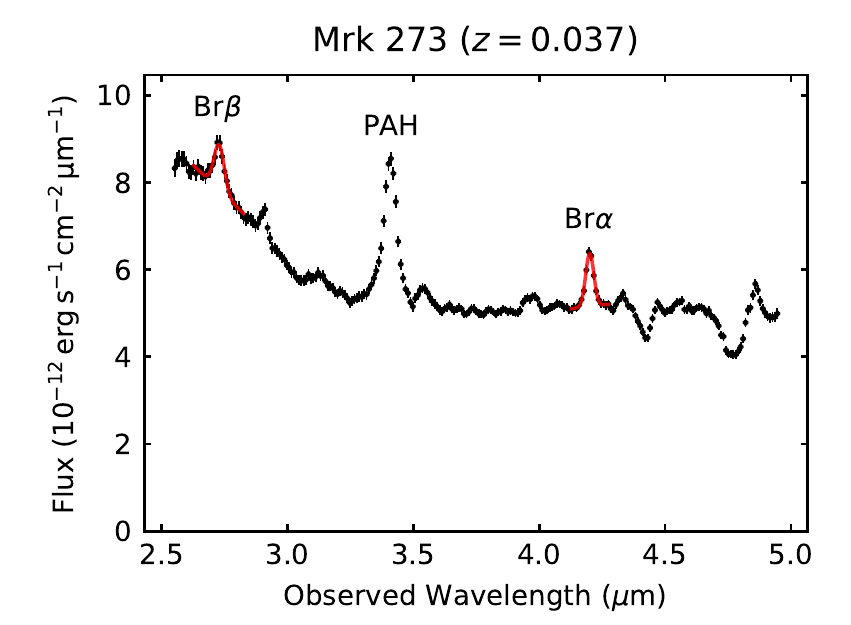}{\columnwidth}{}}
\caption{The 2.5--5.0~$\mu$m near-infrared spectra of galaxies that show
Br$\beta$/Br$\alpha$ line ratios significantly higher than that for case~B.
The best-fit Gaussian profiles for the Br$\alpha$ and Br$\beta$ lines
are plotted as red curves.
\label{fig:anospe}}
\end{figure}

\begin{figure*}
\gridline{\fig{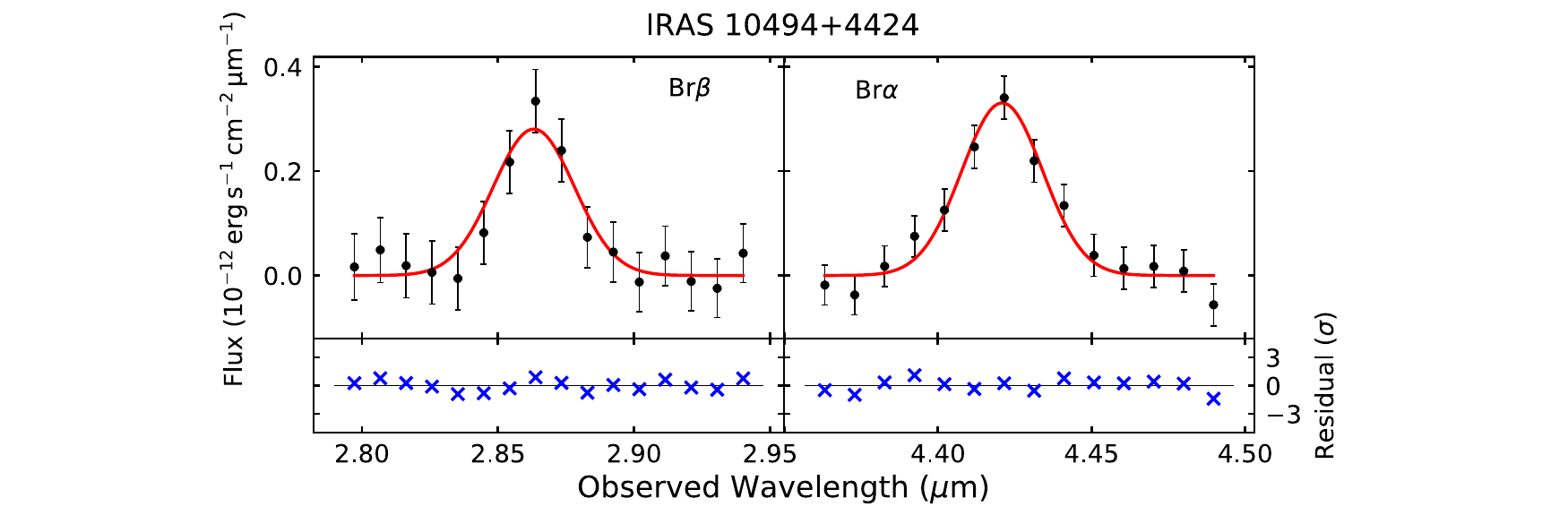}{\textwidth}{}}
\gridline{\fig{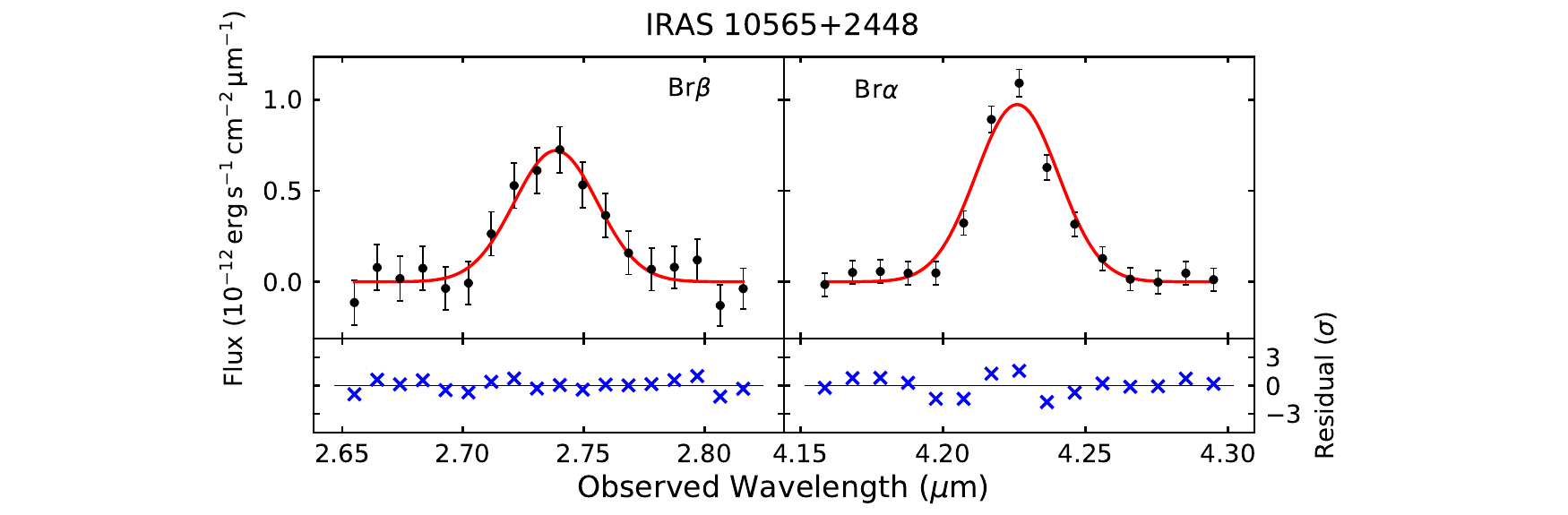}{\textwidth}{}}
\gridline{\fig{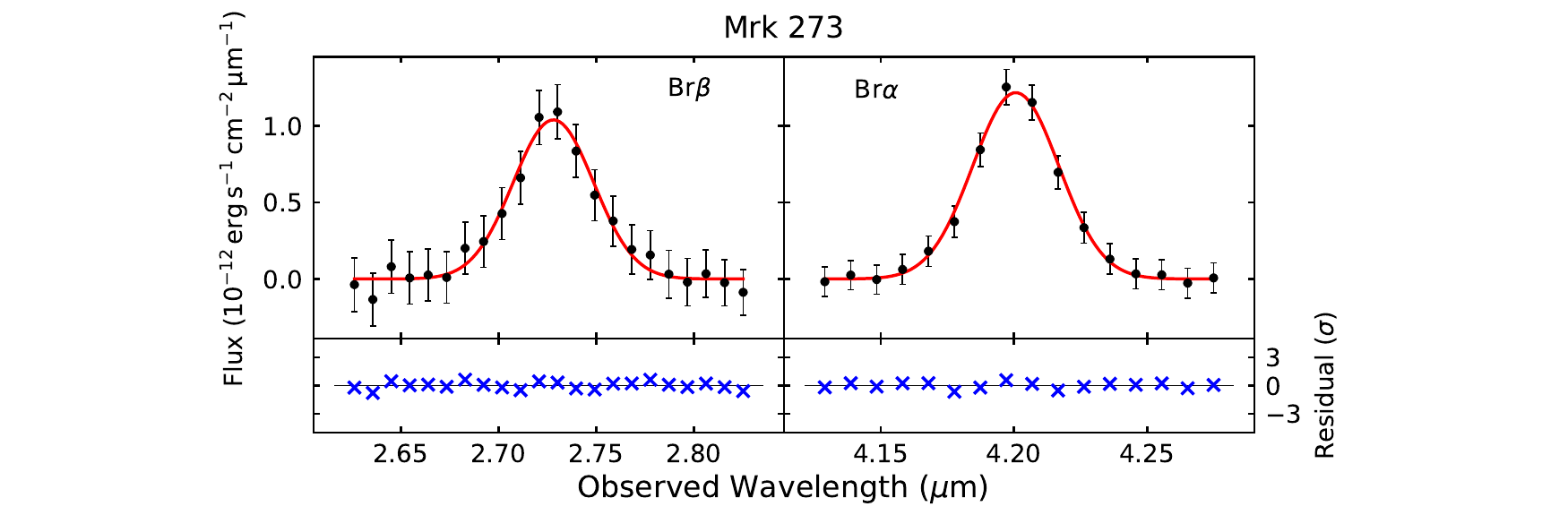}{\textwidth}{}}
\caption{Spectra around the Br$\alpha$ (right) and Br$\beta$ (left) lines
for the three galaxies that show
anomalous Br$\beta$/Br$\alpha$ line ratios.
The underlying continuum has been subtracted.
The best-fit Gaussian profile
is plotted as a solid red curve.
The residuals from the best fit are also displayed
in the bottom panels using blue crosses.
\label{fig:anobr}}
\end{figure*}

We examined the spectra of the four galaxies
that deviate by more than 3$\sigma$ from case~B in Figure~\ref{fig:brflux};
they are IRAS 04074$-$2801, IRAS 10494$+$4424,
IRAS 10565$+$2448, and Mrk~273.
Among them, IRAS~04074$-$2801 is relatively faint,
and the continuum is considerably affected by a fringe-like pattern.
We found that if the wavelength range for fitting the Br$\beta$ line
is widened by a factor of 1.5,
the Br$\beta$ line flux decreases by $\sim20\%$ for this galaxy.
Thus, we excluded IRAS~04074$-$2801
from our discussion of the high Br$\beta$/Br$\alpha$ line ratio
because of the large uncertainties in
determining the continuum underlying the line.
For the remaining three galaxies
for which the spectra are shown in Figure~\ref{fig:anospe},
a change in the wavelength range does not affect the line fluxes
by more than 5\%.
We conclude that the three galaxies (shown as blue circles in Figure~\ref{fig:brflux}),
IRAS~10494$+$4424,
IRAS~10565$+$2448, and Mrk~273,
show Br$\beta$/Br$\alpha$ line ratios of 
0.96$\pm$0.12,
0.883$\pm$0.085, and 
1.086$\pm$0.053,
respectively,
all significantly higher than the case B ratio (0.565).
These line-ratio anomalies are not explainable
with case B theory and dust extinction,
which could reduce but not increase the Br$\beta$/Br$\alpha$ line ratio.
We show the spectra around the Br$\alpha$ and
Br$\beta$ lines for these three galaxies in Figure~\ref{fig:anobr}.

We found no distinct physical properties to
distinguish the three galaxies (IRAS~10494$+$4424,
IRAS~10565$+$2448, and Mrk~273)
with high Br$\beta$/Br$\alpha$ line ratios
from the other sources.
The optical classifications of these galaxies are
LINER, \ion{H}{2} galaxy, and Seyfert~2 \citep{Veilleux1999osi}.
The three galaxies have been observed
at infrared wavelengths, as reported in several publications
\citep[e.g.,][]{Imanishi2008si2,Veilleux2009sqa,Lee2012ani}.
Infrared properties, such as the strengths of their PAH emissions,
were compared to those of other ULIRGs,
but no significant differences were reported.

One common observational property
is that the three objects have relatively low redshifts
($z\sim0.09$ for IRAS 10494$+$4424
and $z\sim0.04$ for IRAS~10565$+$2448 and Mrk 273)
compared to the others in our sample;
accordingly, the Br$\alpha$ and Br$\beta$ lines
are detected with a high S/N ratio,
and the Br$\beta$/Br$\alpha$ line ratio
is well-determined in the three galaxies.
Thus, they provide
clear detections of deviations of the Br$\beta$/Br$\alpha$ line
ratio from case B.
This implies that
anomalous Br$\beta$/Br$\alpha$ line ratios
might also exist in faint galaxies for which we have not been able
to verify its presence
because of large uncertainties in the observed Br$\beta$/Br$\alpha$ line ratios.

\begin{figure}
\plotone{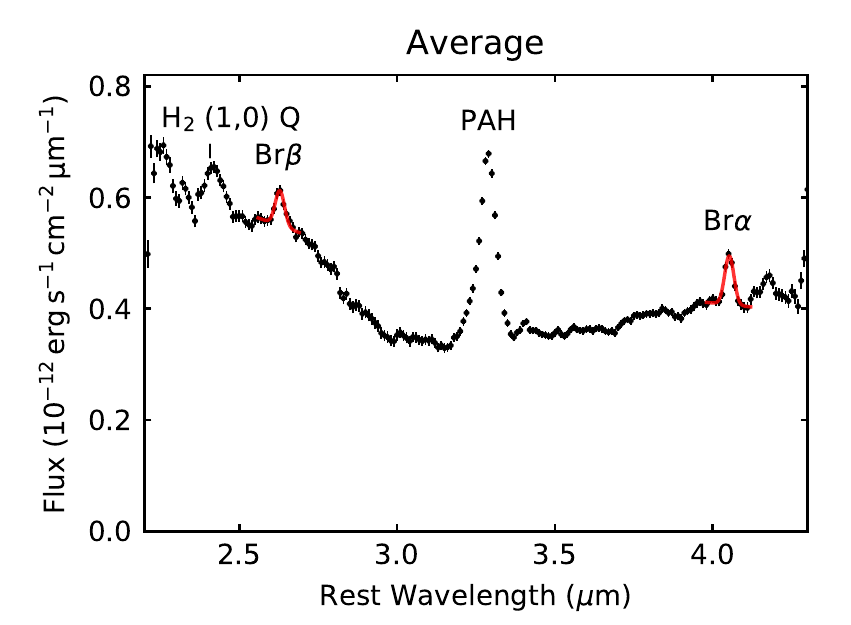}
\caption{Averaged near-infrared spectrum of
35 galaxies with $F_{\mathrm{Br}\alpha}<10^{-14}$~erg~s$^{-1}$~cm$^{-2}$.
The best-fit Gaussian profiles for the Br$\alpha$ and Br$\beta$ lines
are plotted as red curves.
\label{fig:avespe}}
\end{figure}

To investigate whether the anomaly is actually found in faint galaxies,
we averaged the near-infrared spectra of 35 galaxies with
$F_{\mathrm{Br}\alpha}<10^{-14}$~erg~s$^{-1}$~cm$^{-2}$.
Each spectrum was corrected for redshift and
was averaged in rest wavelength.
The averaged spectrum of the 35 galaxies is shown in Figure~\ref{fig:avespe}.
We measured the values of $F_{\mathrm{Br}\alpha}$ and $F_{\mathrm{Br}\beta}$
from the averaged spectrum using Gaussian fitting,
obtaining
$F_{\mathrm{Br}\alpha}=(3.21\pm0.20)\times10^{-15}$~erg~s$^{-1}$~cm$^{-2}$
and $F_{\mathrm{Br}\beta}=(2.30\pm0.16)\times10^{-15}$~erg~s$^{-1}$~cm$^{-2}$.
This yields a Br$\beta$/Br$\alpha$ line ratio of $0.716\pm0.065$,
which is 2.3$\sigma$ higher than the case B value of 0.565.
Although the significance of this result is not high enough ($<3\sigma$),
it is opposite to the expectation that ULIRGs should show high dust extinction;
i.e., the Br$\beta$/Br$\alpha$ line ratio should be lower than 0.565.
We thus conclude that in future high-sensitivity observations,
such as those with \textit{the James Webb Space Telescope (JWST)},
anomalous Br$\beta$/Br$\alpha$ line ratios
may be found in faint galaxies for which we have not been able
to determine the presence of the anomaly
with current \textit{AKARI} observations.

\subsection{Estimate of Contamination of Brackett Lines}

One possible cause of the line-ratio anomaly
is that the Brackett lines may be blended with other features.
The spectral resolution of our observations
is not high ($\sim$0.04--0.05~$\mu$m);
therefore, the apparently high Br$\beta$/Br$\alpha$ line ratio may be caused
by contamination due to other features.
We looked for such features,
with wavelengths close to those of the Brackett lines.
However, there are few observations around the wavelengths of the Brackett lines
(especially around the Br$\beta$ line)
because the wavelength is difficult to access
from the ground due to atmospheric absorption.
Little information is therefore available about
possible contaminant features.

We reviewed a list of infrared atomic lines
provided by \textit{ISO} observations \citep[Table~4.10 of][]{Glass1999h}
and found no candidate lines near the wavelengths of the Br$\beta$ and Br$\alpha$ lines.
Among molecular features \citep[Table~4.10 of][]{Cox2000a},
we found one candidate line, H$_2$~(1,0)~O(2),
for which the wavelength is 2.627~$\mu$m.
The wavelength of this line is very close to that of the Br$\beta$ line;
therefore, the flux of the Br$\beta$ line could be overestimated
due to contamination by the molecular hydrogen line.
We discuss below the possible effect of
molecular hydrogen contamination
on the Br$\beta$/Br$\alpha$ line ratio
for the three galaxies with anomalous Brackett-line line ratios,
IRAS~10494$+$4424, IRAS~10565$+$2448, and Mrk~273.

We estimated the flux of the H$_2$ (1,0) O(2) line
($F_{\mathrm{O}(2)}$)
using another molecular hydrogen rotation-vibration
line, H$_2$ (1,0) O(3), which has a rest wavelength
of 2.802 $\mu$m and which is simultaneously detected in
our observations.
\cite{Black1987fei} calculated
the flux ratios of molecular hydrogen
rotation-vibration lines both for
fluorescence excitation and shock excitation.
In previous ground-based $K$-band observations,
the H$_2$ lines were shown to be
thermally excited at a temperature of $\sim2000$~K
in IRAS~10494$+$4424 \citep{Murphy2001kbs},
IRAS~10565$+$2448, and Mrk~273 \citep{Goldader1997s}.
Thus, we adopt the flux ratio
$F_{\mathrm{O}(2)}=0.26F_{\mathrm{O}(3)}$
from the shock model ($T=2000$ K) of \cite{Black1987fei},
where $F_{\mathrm{O}(3)}$ is the flux of the H$_2$ (1,0) O(3) line.

\begin{figure}
\gridline{\fig{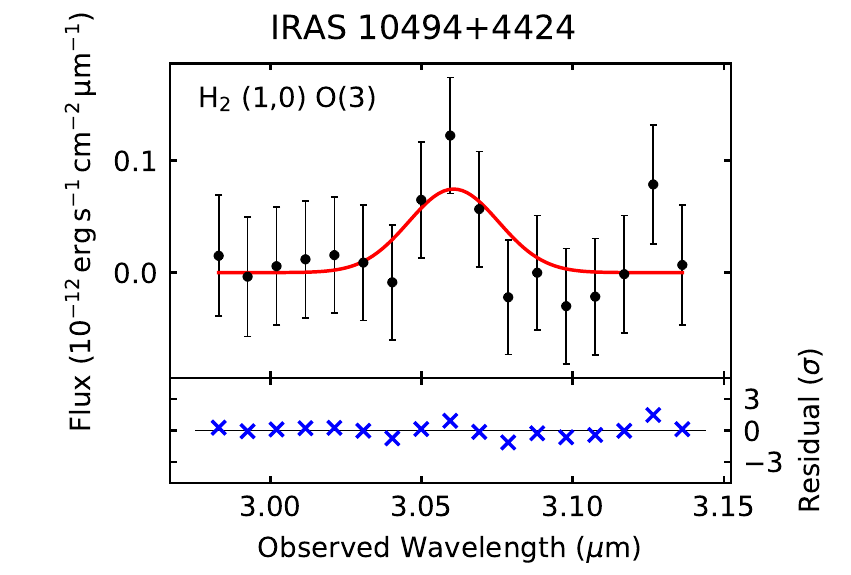}{\columnwidth}{}}
\gridline{\fig{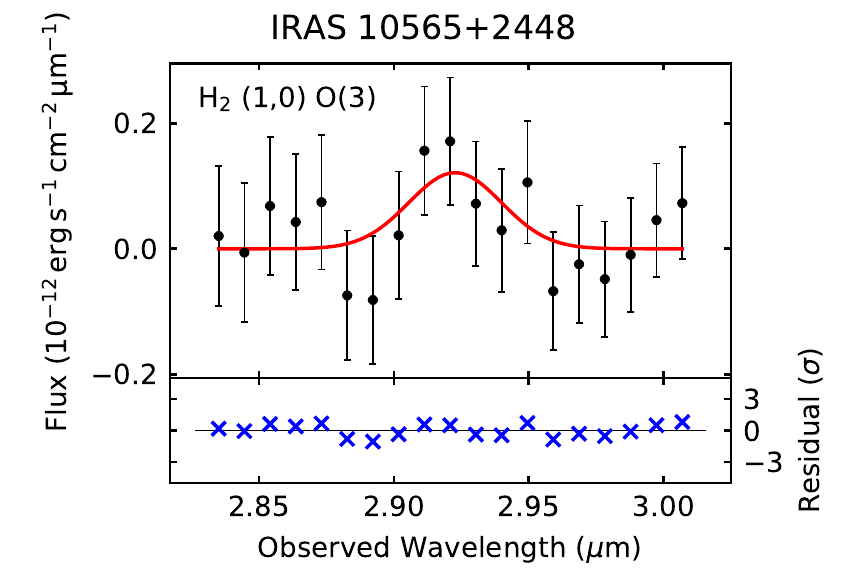}{\columnwidth}{}}
\gridline{\fig{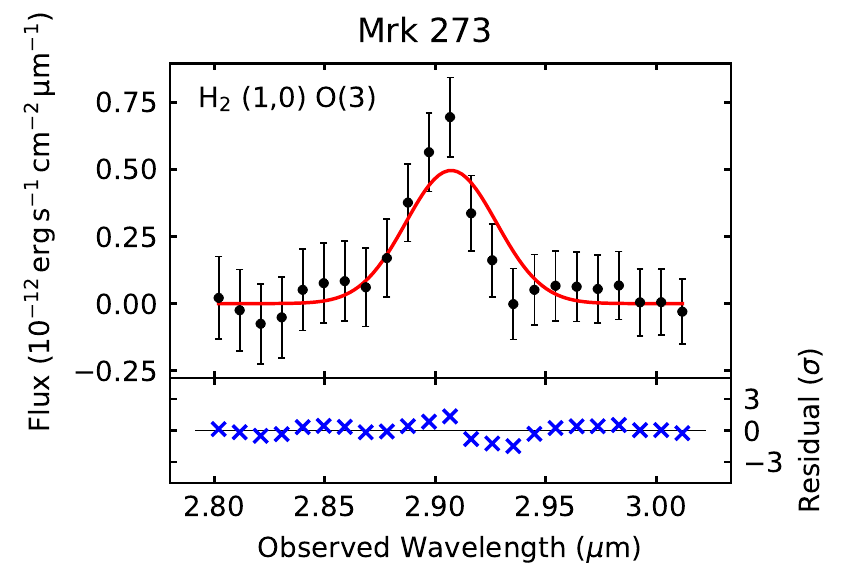}{\columnwidth}{}}
\caption{Spectra around the H$_2$ (1,0) O(3) line.
The underlying continuum has been subtracted.
The best-fit Gaussian profile
is plotted as a solid red curve.
The residuals from the best fit are also displayed
in the bottom panels using blue crosses.
\label{fig:h2spe}}
\end{figure}

We measured $F_{\mathrm{O}(3)}$
for the three galaxies that show
anomalous Brackett-line ratios
using Gaussian fitting.
We fixed the width and central wavelength of the Gaussian profile
and used the normalization of the Gaussian profile and the linear continuum
as free parameters.
Figure~\ref{fig:h2spe} shows the results of the Gaussian fitting
of the H$_2$ (1,0) O(3) line.
We detected the H$_2$ (1,0) O(3) line
with moderate significance ($>2.8\sigma$)
in all three galaxies.
The measured flux is summarized in Table~\ref{tab:h2flux},
along with the 1$\sigma$ statistical error.
The 1$\sigma$ systematic error is estimated to be $\sim10\%$
of the flux.

\begin{deluxetable*}{cccc}
\tablecaption{Correction for Contamination by Molecular Hydrogen Line \label{tab:h2flux}}
\tablewidth{0pt}
\tablehead{\colhead{}&\colhead{(Observed)}&\colhead{(Predicted)}&\colhead{}\\
\colhead{Object}&\colhead{$F_{\mathrm{O}(3)}$}&\colhead{$F_{\mathrm{O}(2)}$\tablenotemark{a}}&\colhead{$F_{\mathrm{Br}\beta}^\mathrm{cor}/F_{\mathrm{Br}\alpha}$\tablenotemark{b}}\\
\colhead{Name}&\multicolumn{2}{c}{($10^{-15}$~erg~s$^{-1}\;$cm$^{-2}$)}&\colhead{}
}
\startdata
IRAS~10494$+$4424 & $ 2.79\pm0.92$ & $0.72\pm0.24$ & $0.89 \pm0.12 $ \\
IRAS~10565$+$2448 & $ 5.2 \pm1.9 $ & $1.36\pm0.49$ & $0.845\pm0.085$ \\
Mrk~273           & $25.6 \pm3.0 $ & $6.66\pm0.79$ & $0.952\pm0.054$ \\
\enddata
\tablenotetext{a}{Flux of the H$_2$ (1,0) O(2) line
derived from the H$_2$ (1,0) O(3) line
assuming $F_{\mathrm{O}(2)}=0.26F_{\mathrm{O}(3)}$
\citep[2000~K shock model of][]{Black1987fei}.}
\tablenotetext{b}{Br$\beta$/Br$\alpha$ line ratio corrected
for contamination by the molecular hydrogen line.
The predicted flux of the H$_2$ (1,0) O(2) line
is subtracted from the observed flux of the Br$\beta$ line.}
\end{deluxetable*}

Using the H$_2$ (1,0) O(3) line flux, 
we determined the H$_2$ (1,0) O(2) line flux,
subtracted it from the Br$\beta$ line flux, and
calculated the flux ratio of the pure Br$\beta$ line
to the Br$\alpha$ line.
We summarize the results
in Table~\ref{tab:h2flux}
($F_{\mathrm{Br}\beta}^\mathrm{cor}/F_{\mathrm{Br}\alpha}$).
The ratio is still more than 3$\sigma$ higher than the case~B value of 0.565
for IRAS~10565$+$2448 and Mrk~273.
Thus, contamination of the Br$\beta$ line
does not provide a full explanation of
the anomalous line ratio
at least for these galaxies.
For IRAS~10494$+$4424, the significance of the anomaly is 2.7$\sigma$
after the correction of the H$_2$ line.
Hence, the detection of an anomalous line ratio in this ULIRG is
admittedly not as robust as those in the other two,
but it is still moderately significant.
We thus retain this galaxy in the discussion below.
We conclude that the anomalous Br$\beta$/Br$\alpha$ ratio is real
and that conditions intrinsic to the ionized gas itself make the ratio
anomalous.

In summary, for three out of 33 ULIRGs
wherein we detected both the Br$\alpha$ and Br$\beta$ lines,
we found anomalous Br$\beta$/Br$\alpha$ line ratios.
The ratios are significantly higher than the case~B value
even after the subtraction of possible contamination of the H$_2$ (1,0) O(2) line,
at least in two sources,
and are not explained by the effects of dust extinction.
We also found that ULIRGs have a tendency to
exhibit high Br$\beta$/Br$\alpha$ line ratios.
As we discuss in Appendix~\ref{sec:gal},
the case~B line ratio explains well the Br$\beta$/Br$\alpha$ line ratio
in Galactic \ion{H}{2} regions.
This indicates that conditions in the \ion{H}{2} regions
in those ULIRGs with a Br$\beta$/Br$\alpha$ anomaly are entirely different
from the conditions in Galactic \ion{H}{2} regions.

\section{Interpretation of the Anomaly}
\label{sec:ioa}

In this section, we discuss
how to interpret the high Br$\beta$/Br$\alpha$ ratio.
To investigate the \ion{H}{1} line ratios,
we first consider how the level populations
of the hydrogen atoms are determined by assuming
three possible excitation mechanisms:
recombination, collisional excitation,
and resonant excitation.
Then, we discuss possible explanations
for the high Br$\beta$/Br$\alpha$ ratio
separately for optically thin and optically thick cases.
In this section, we use the Einstein coefficients
from \cite{Johnson1972a} and recombination coefficients
from \cite{Verner1996a}.

\subsection{Excitation Mechanisms for the Hydrogen Atoms}
\label{sec:mech}

\subsubsection{Recombination}
\label{sec:reco}

As the first excitation mechanism,
we discuss the recombination process.
In the low-density limit,
wherein collisions are negligible,
the hydrogen level populations are
determined by recombination and radiative transitions.
In this case,
the equilibrium equation
for the level population of a state with
principal quantum number $\mathcal{N}$
can be written as
\begin{equation}
n_\mathrm{p}n_\mathrm{e}\alpha_\mathcal{N}
+\sum^\infty_{\mathcal{N}'=\mathcal{N}+1}n_{\mathcal{N}'}A_{\mathcal{N}',\mathcal{N}}
=n_\mathcal{N}\sum^{\mathcal{N}-1}_{\mathcal{N}''=2}A_{\mathcal{N},\mathcal{N}''},
\label{eq:neq}
\end{equation}
where $n_\mathrm{p}$, $n_\mathrm{e}$, and $n_\mathcal{N}$
are the number densities of protons, electrons, and
hydrogen atoms in quantum state $\mathcal{N}$, respectively;
$\alpha_\mathcal{N}$ is the recombination coefficient for level $\mathcal{N}$;
and $A_{\mathcal{N}',\mathcal{N}}$ is the Einstein A coefficient
for the $\mathcal{N}'\rightarrow \mathcal{N}$ transition.
We assume case B conditions \citep{Osterbrock2006agn},
for which the Lyman lines are taken to be optically thick,
so that the summation on the right-hand side ends at $\mathcal{N}''=2$.

Using the cascade matrix $C_{\mathcal{N}',\mathcal{N}}$,
which is the probability that
a population in level $\mathcal{N}'$ undergoes
a transition to level $\mathcal{N}$ via all possible routes \citep{Seaton1959s},
we can rewrite Equation~(\ref{eq:neq}) as
\begin{equation}
n_\mathcal{N}A_\mathcal{N}=
n_\mathrm{p}n_\mathrm{e}\sum^\infty_{\mathcal{N}'=\mathcal{N}}
\alpha_{\mathcal{N}'}C_{\mathcal{N}',\mathcal{N}},
\label{eq:nstate}
\end{equation}
where we have written $\sum^{\mathcal{N}-1}_{\mathcal{N}''=2}A_{\mathcal{N},\mathcal{N}''}=A_\mathcal{N}$.
Thus, we have
\begin{equation}
\frac{n_{\mathcal{N}}}{n_{\mathcal{N}'}}=\frac{A_{\mathcal{N}'}}{A_{\mathcal{N}}}
\frac{\sum^\infty_{l={\mathcal{N}}}\alpha_{l}C_{l,{\mathcal{N}}}}
{\sum^\infty_{l'={\mathcal{N}'}}\alpha_{l'}C_{l',{\mathcal{N}'}}}
\label{eq:n6n5}
\end{equation}
for the ratio of the level populations of states
$\mathcal{N}$ and $\mathcal{N}'$.
The cascade matrix can be written in terms of the Einstein coefficients,
which do not depend on gas properties such as temperature.
The recombination coefficients depend weakly on temperature;
however, in Equation~(\ref{eq:n6n5}), this dependence is almost canceled out
because we have the coefficients both in the numerator and the denominator.
Thus, the level population depends only weakly on temperature
for the case in which it is determined by the recombination process.

\subsubsection{Collisional Excitation}

Next,
we consider the collisional excitation mechanism.
In the high-density limit, wherein
the level population is entirely determined by collisions,
the level population reaches thermal equilibrium
and follows the Boltzmann distribution.
The ratio of level populations for this case can be written as
\begin{equation}
\frac{n_\mathcal{N}}{n_{\mathcal{N}'}}=\frac{g_\mathcal{N}}{g_{\mathcal{N}'}}
\frac{\mathrm{exp}\left( -\frac{E_\mathcal{N}}{kT}\right)}{\mathrm{exp}\left(-\frac{E_{\mathcal{N}'}}{kT}\right)},
\label{eq:kt}
\end{equation}
where $g_\mathcal{N}$ is the degeneracy of state $\mathcal{N}$,
$k$ is Boltzmann's constant,
$E_\mathcal{N}$ is the energy of state $\mathcal{N}$
relative to the ground state, and $T$ is the gas temperature.
The critical densities for transitions
$\mathcal{N}=1\rightarrow5$ and 6 are
on the order of $\sim10^{11}$~cm$^{-3}$ \citep{Storey1995rli};
therefore, gas densities higher than this threshold
are required to achieve thermal equilibrium for the levels
related to the Br$\alpha$ and Br$\beta$ line emissions.

\subsubsection{Resonant Excitation}

The third excitation mechanism is
resonant excitation.
One example wherein this process becomes important
is the Bowen resonance of \ion{O}{3} lines \citep{Osterbrock2006agn}.
The wavelength of the \ion{O}{3} $2p^2\ {}^3P_2$ -- $3d^3\ {}^3P_2^o$
line (303.80 \AA) is accidentally coincident with that of
the \ion{He}{2} L$\alpha$ line (303.78 \AA);
therefore, the $3d^3\ {}^3P_2^o$ level of \ion{O}{3} is
pumped by the \ion{He}{2} L$\alpha$ line.
This results in an enhancement of the \ion{O}{3} lines
originating from the $3d^3\ {}^3P_2^o$ state.
A similar situation would occur for hydrogen lines
if a line were to exist with a wavelength
close to that of the $\mathcal{N}=\mathcal{N}'\rightarrow1$
transitions, i.e., the Lyman series.
In this case, the line will pump the electrons
from the hydrogen ground state
to state $\mathcal{N}'$,
resulting in an enhancement
of the level population of the $\mathcal{N}'$ state.

Based on the three excitation mechanisms described above,
we consider some possible explanations for the high
Br$\beta$/Br$\alpha$ line ratios.

\subsection{Optically Thin Case}
\label{sec:thin}

We first consider the case in which
the Brackett lines are optically thin.
In this case, once the level population of neutral hydrogen
is determined, the flux ratios of the \ion{H}{1} lines
are fixed.
The emergent Br$\beta$/Br$\alpha$ line ratio
is then
\begin{equation}
\frac{F_{\mathrm{Br}\beta}}{F_{\mathrm{Br}\alpha}}
=\frac{n_6}{n_5}\frac{A_{\mathrm{Br}\beta}}{A_{\mathrm{Br}\alpha}}
\frac{h\nu_{\mathrm{Br}\beta}}{h\nu_{\mathrm{Br}\alpha}}
\sim0.440\frac{n_6}{n_5},
\label{eq:f6f5}
\end{equation}
where $n_6/n_5$ is the ratio of the populations of hydrogen atoms in levels
$\mathcal{N}=6$ and $\mathcal{N}=5$,
$A_\mathrm{line}$ is the Einstein A coefficient for the line,
$h$ is Planck's constant, and $\nu_\mathrm{line}$ is
the frequency of the line.
This equation shows that
in order to explain a high Br$\beta$/Br$\alpha$ line ratio,
the level population in $\mathcal{N}=6$ must be enhanced
relative to that in $\mathcal{N}=5$.

At low densities, where the recombination process is dominant,
the level populations are determined by Equation~(\ref{eq:n6n5}).
Assuming $T=10000$~K as the gas temperature,
we obtain $F_{\mathrm{Br}\beta}/F_{\mathrm{Br}\alpha}=0.565$,
i.e., the case B ratio,
by substituting Equation~(\ref{eq:n6n5}) into
Equation~(\ref{eq:f6f5}).
In contrast, in the high-density limit
where the collisional process is dominant,
the hydrogen level populations are determined by Equation~(\ref{eq:kt}).
Again, assuming $T=10000$~K as the gas temperature,
we obtain $F_{\mathrm{Br}\beta}/F_{\mathrm{Br}\alpha}=0.522$
by substituting Equation~(\ref{eq:kt}) into
Equation~(\ref{eq:f6f5}).
If we take the limit $T\rightarrow\infty$,
Equation~(\ref{eq:kt}) gives
$n_6/n_5=g_6/g_5=1.44$ as the high-$T$ limit,
and we obtain $F_{\mathrm{Br}\beta}/F_{\mathrm{Br}\alpha}=0.634$.
This is the highest line ratio achievable with collisional excitation but it is
still lower than the observed values.
Thus, we cannot explain the high Br$\beta$/Br$\alpha$ line ratios
in either the high-density or the low-density limits.

\begin{figure}
\plotone{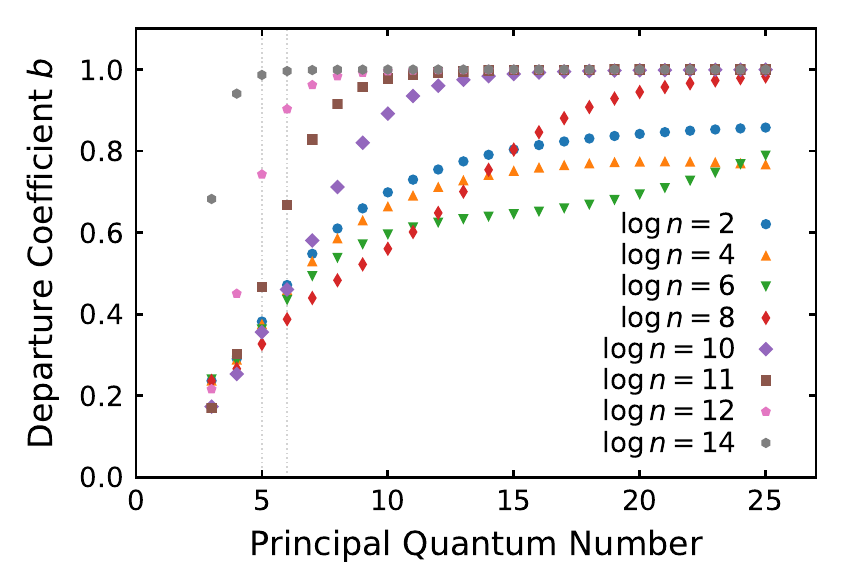}
\caption{Departure coefficients for hydrogen atoms in states
$\mathcal{N}\leq25$ for
various total hydrogen densities $n$
at $T=10000$~K assuming case B conditions
\citep[data taken from][]{Storey1995rli}.
The levels $\mathcal{N}=5$ and 6 are denoted with gray dotted lines.
\label{fig:depcoef}}
\end{figure}

In general,
the level populations are affected
both by the recombination and collisional process,
so the combined results are expected to lie
somewhere between the above two limits.
This is expressed in terms of the departure coefficient $b_\mathcal{N}$,
which is the fractional departure of the population of state $\mathcal{N}$ from that
in thermal equilibrium, $n_\mathcal{N}^\mathrm{th}$;
i.e., $n_\mathcal{N}=b_\mathcal{N}n_\mathcal{N}^\mathrm{th}$.
\cite{Storey1995rli} calculated the $b_\mathcal{N}$ coefficients
for several gas densities,
and we show their results in Figure~\ref{fig:depcoef}.
Hereafter, we denote
the total hydrogen number density by $n$;
i.e., we write $n(\mathrm{H}^0)+n(\mathrm{H}^+)=n$,
where $n(\mathrm{H}^0)$ and $n(\mathrm{H}^+)$
are the number densities of neutral and ionized hydrogen, respectively.
In the ionized gas, we assume that the hydrogen atoms are fully ionized,
so that $n_\mathrm{e}\sim n$.
The level populations for $\mathcal{N}=6$ and $\mathcal{N}=5$
are affected by the collisional process
for densities higher than the critical density $n\sim10^{11}$~cm$^{-3}$.

\begin{figure}
\plotone{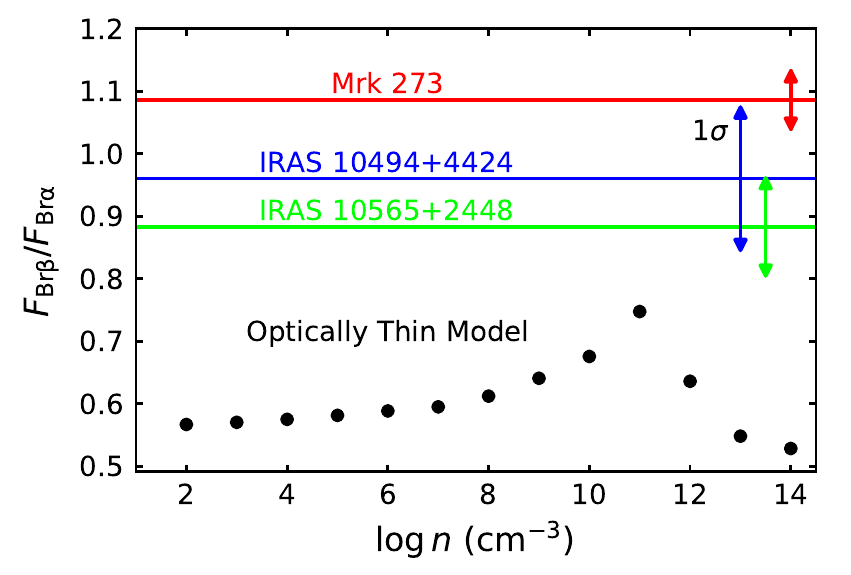}
\caption{The $F_{\mathrm{Br}\beta}/F_{\mathrm{Br}\alpha}$ ratio
for the optically thin case, for various total hydrogen densities $n$.
The horizontal red, green, and  blue lines show the observed $F_{\mathrm{Br}\beta}/F_{\mathrm{Br}\alpha}$
ratios for Mrk~273, IRAS~10565$+$2448, and IRAS~10494$+$4424, respectively.
The 1$\sigma$ uncertainty ranges are shown by the vertical arrows.
\label{fig:thin}}
\end{figure}

Using the $b_\mathcal{N}$ coefficients, we can determine
the $n_6/n_5$ ratio for each gas density
and so derive the Br$\beta$/Br$\alpha$ line
ratio from Equation~(\ref{eq:f6f5}).
We show these results in Figure~\ref{fig:thin}.
At low densities, the Br$\beta$/Br$\alpha$ line ratio
is consistent with the case B ratio.
When the density becomes $n\ge10^{10}$~cm$^{-3}$,
collisional excitation starts to contribute
to the $\mathcal{N}=6$ state.
This causes an enhancement of the Br$\beta$ line.
At densities higher than $n\geq10^{12}$~cm$^{-3}$,
the $\mathcal{N}=5$ state also begins to be collisionally excited,
and so the Br$\beta$/Br$\alpha$ line ratio
approaches the ratio in thermal equilibrium.

Figure~\ref{fig:thin} indicates that the Br$\beta$/Br$\alpha$ line
ratio becomes as high as $\sim$0.75
at $n\sim10^{11}$~cm$^{-3}$.
This ratio is within 1.6$\sigma$ and 1.8$\sigma$ of the observed ratios for
IRAS 10565$+$2448 (0.88$\pm$0.09) and IRAS 10494$+$4424 (0.96$\pm$0.12), respectively,
but it is still more than 3$\sigma$ lower than that observed in Mrk~273 (1.09$\pm$0.05).
Thus, we conclude that we cannot explain the anomaly
with just the recombination and collisional processes
in the optically thin case.

With resonant excitation,
the $\mathcal{N}=6$ state is enhanced
if a strong line with a wavelength
comparable to that of the transition
$\mathcal{N}=6\rightarrow1$ (937.8~\AA) exists.
As we discuss in detail in Appendix~\ref{sec:ap6},
we found that if this (unknown) line
have a transition probability
comparable to those of forbidden lines,
then the resonant process
would be able to make the Br$\beta$/Br$\alpha$ line
ratio anomalously high.

We reviewed the atomic and molecular data
currently available to search for possible resonant lines.
For the line data, we used the line list
provided in the Cloudy program \citep{Ferland1998c9n}.
Cloudy is a spectral-synthesis code designed to numerically simulate
an astrophysical plasma and its emissions.
Extensive atomic and molecular data are
collected in the code
(references are available in a file
distributed along with the code).
We searched for possible resonant lines with
a wavelength of $\sim$937.8~\AA\ 
within a velocity range of $\sim10$~km~s$^{-1}$,
corresponding to the thermal velocity at $T=10000$~K \citep{Osterbrock2006agn}.
We found no candidates in the Cloudy data,
and thus we exclude this resonant process
as a possible cause of the anomalous
Br$\beta$/Br$\alpha$ line ratio.

Based on the above discussion,
we conclude that we cannot explain
the anomalous Br$\beta$/Br$\alpha$ line ratio
if the Brackett lines are optically thin.

\subsection{Optically Thick Case}

The alternative is that the Brackett lines
are optically thick
and the observed line ratio
deviates from Equation~(\ref{eq:f6f5}).
In this case, it is possible to explain the high Br$\beta$/Br$\alpha$ line ratio
if the Br$\alpha$ line becomes optically thick
and saturates, whereas the Br$\beta$ line
remains optically thin.
Herein, we discuss the optical-depth effect
on this ratio.

\subsubsection{Optical Depth of Brackett Lines}

In a uniform gas,
the optical depth $\tau_{\mathcal{N}',\mathcal{N}}$
at the line center for the transition $\mathcal{N}'\rightarrow \mathcal{N}$
is given as
\begin{equation}
\tau_{\mathcal{N}',\mathcal{N}}=\int_0^R \alpha_{\mathcal{N},\mathcal{N}'}
n_\mathcal{N} \mathrm{d} l \sim \alpha_{\mathcal{N},\mathcal{N}'}N_\mathcal{N},
\label{eq:tau}
\end{equation}
where $R$ is the size of the gas, $\alpha_{\mathcal{N},\mathcal{N}'}$ is the absorption cross-section
of the transition $\mathcal{N}\rightarrow \mathcal{N}'$,
$n_\mathcal{N}$ is the number density of neutral hydrogen in state $\mathcal{N}$,
and $N_\mathcal{N}$ is the column density of neutral hydrogen in state $\mathcal{N}$.
The optical depth of the Brackett lines
is proportional to the column density
of neutral hydrogen in the quantum state $\mathcal{N}=4$.

Assuming a Gaussian profile as a line velocity profile,
the absorption cross-section is related to the Einstein B coefficient, $B_{\mathcal{N},\mathcal{N}'}$, by
\begin{equation}
\alpha_{\mathcal{N},\mathcal{N}'}=\frac{hc}{4\pi^{3/2}}
\frac{B_{\mathcal{N},\mathcal{N}'}}{v_\mathrm{Dop}},
\label{eq:al}
\end{equation}
where $c$ is the speed of light, and $v_\mathrm{Dop}$
is the Doppler velocity half width,
the distance from line center
where the line profile falls to $\mathrm{e}^{-1}$ of its peak.
If the line profile
is determined solely by thermal motions,
the Doppler width can be written
as $v_\mathrm{Dop}=v_\mathrm{Therm}=\sqrt{2kT/m_\mathrm{H}}$,
where $m_\mathrm{H}$ is the mass of a hydrogen atom.
At $T=10000$~K,
we have $v_\mathrm{Therm}\sim13$~km~s$^{-1}$.
If a turbulent motion with a velocity $v_\mathrm{Turb}$ affects the line width,
$v_\mathrm{Dop}=\sqrt{v_\mathrm{Therm}^2+v_\mathrm{Turb}^2}$.

Substituting Equation~(\ref{eq:al}) into
Equation~(\ref{eq:tau})
and assuming $T=10000$~K,
we obtain the line optical depths of
the Br$\alpha$ and Br$\beta$ lines as
\begin{eqnarray}
\tau_{\mathrm{Br}\alpha}&\sim&1.0
\left(\frac{N_4}{1.6\times10^{11}\ \mathrm{cm}^{-2}}\right)
\left(\frac{v_\mathrm{Dop}}{10\ \mathrm{km}~\mathrm{s}^{-1}}\right)^{-1},\label{eq:taua}\\
\tau_{\mathrm{Br}\beta}&\sim&0.11
\left(\frac{N_4}{1.6\times10^{11}\ \mathrm{cm}^{-2}}\right)
\left(\frac{v_\mathrm{Dop}}{10\ \mathrm{km}~\mathrm{s}^{-1}}\right)^{-1}.\label{eq:taub}
\end{eqnarray}
Thus, assuming $v_\mathrm{Dop}\sim10$~km~s$^{-1}$, for instance,
we find that the Br$\alpha$ line becomes optically thick
while the Br$\beta$ line is still optically thin
when $N_4\sim2\times 10^{11}$~cm$^{-2}$.
We further discuss the possible conditions that
can produce a high Br$\beta$/Br$\alpha$ line ratio
on the basis of Equations~(\ref{eq:taua}) and (\ref{eq:taub}).

\subsubsection{Possible Conditions Producing a High Brackett-Line Ratio}
\label{sec:pcp}

Herein, we assume that the high Br$\beta$/Br$\alpha$ line ratio
is produced within a single isolated \ion{H}{2} region
ionized by a single star
and that what we observe is an ensemble of such \ion{H}{2} regions.
Within a single \ion{H}{2} region,
we assume that $v_\mathrm{Dop}$ is
determined only by the thermal width,
with $v_\mathrm{Therm}\sim10$~km~s$^{-1}$,
which is a typical velocity observed in nearby \ion{H}{2} regions
\citep[e.g.,][]{Arthur2016t}.
We assume that the Br$\alpha$ line becomes optically thick
within each \ion{H}{2} region.

For a spherical and uniform \ion{H}{2} region,
we next discuss how to make the Br$\alpha$ line optically thick
using the three excitation mechanisms described in \S\ref{sec:mech}.
First, we consider the case in which the recombination process is dominant.
In this case, $n_4$ is determined by Equation~(\ref{eq:nstate}),
which yields $n_4\sim3.6\times10^{-21}(n/\mathrm{cm}^{-3})^2$~cm$^{-3}$.
We can thus write $N_4$ in terms of $n$ and $R$ as
\begin{eqnarray}
N_4&=&n_4R \nonumber\\
&\sim&3.6\times10^{-21}\left(\frac{n}{\mathrm{cm}^{-3}}\right)^2
\left(\frac{R}{\mathrm{cm}}\right)\ \mathrm{cm}^{-2}.
\end{eqnarray}
If we write $nR=N$, where $N$ is the total hydrogen column density, then we have
\begin{equation}
N_4=3.6\times10^{-21}\left(\frac{n}{\mathrm{cm}^{-3}}\right)
\left(\frac{N}{\mathrm{cm}^{-2}}\right)\ \mathrm{cm}^{-2}.\label{eq:col4}
\end{equation}
Thus $N_4$ is proportional to both $n$ and $N$.
If the \ion{H}{2} region is
ionized by a central star that emits the number of ionizing photons per unit time $Q(\mathrm{H})$,
then ionization-equilibrium at $T=10000$~K yields
\begin{eqnarray}
&&Q(\mathrm{H})=\int \alpha_\mathrm{B} n_\mathrm{e}n_\mathrm{p}\mathrm{d} V
\sim \frac{4\pi}{3}\alpha_\mathrm{B} n^2R^3=\frac{4\pi}{3}\alpha_\mathrm{B} N^3n^{-1},\nonumber\\
&\therefore&\quad\!
N=3.37\times10^{20}\left(\frac{Q(\mathrm{H})}{10^{49}\ \mathrm{s}^{-1}}\right)^{\frac{1}{3}}
\left(\frac{n}{\mathrm{cm}^{-3}}\right)^{\frac{1}{3}} \mathrm{cm}^{-2},\label{eq:qnn}
\end{eqnarray}
where $\alpha_\mathrm{B}$ is the total recombination coefficient
for hydrogen in case B, and $Q(\mathrm{H})=10^{49}$~s$^{-1}$
is a typical value for an O star \citep{Osterbrock2006agn}.
Substituting Equation~(\ref{eq:qnn}) into Equation~(\ref{eq:col4}),
we obtain $N_4$ in terms of $n$ and $Q(\mathrm{H})$;
\begin{eqnarray}
N_4&=&2.35\times10^{11} \nonumber\\
&&\times\left(\frac{Q(\mathrm{H})}{10^{49}\ \mathrm{s}^{-1}}\right)^{\frac{1}{3}}
\left(\frac{n}{10^{8}\ \mathrm{cm}^{-3}}\right)^{\frac{4}{3}}\ \mathrm{cm}^{-2}.
\label{eq:n4qn}
\end{eqnarray}
Thus, within a single \ion{H}{2} region with an ionizing source emitting
$Q(\mathrm{H})\sim10^{49}$~s$^{-1}$,
a gas density as high as $n\sim10^8$~cm$^{-3}$ is required to
achieve a column density $N_4\sim2\times10^{11}$~cm$^{-2}$
that is large enough to make the Br$\alpha$ line optically thick.

At a density $n=10^8$~cm$^{-3}$,
the collisional process is not dominant
as the excitation mechanism
for the population in the quantum state $\mathcal{N}=4$
because the critical density for the $\mathcal{N}=1\rightarrow4$
transition is $\sim$10$^{12}$~cm$^{-3}$ \citep{Storey1995rli}.
This is also shown in Figure~\ref{fig:depcoef}.
The relative difference between the $b_4$ coefficients at $n=10^8$~cm$^{-3}$ 
and at $10^2$~cm$^{-3}$ is less than $10\%$,
indicating that the $\mathcal{N}=4$ state
is not dominantly populated by the collisional process
at densities $n\leq10^8$~cm$^{-3}$.
Thus, Equation~(\ref{eq:n4qn}),
in which only the recombination process is considered,
is valid if we take collisional excitation into account
at a density of $n=10^8$~cm$^{-3}$.

Resonant excitation would enhance
the $\mathcal{N}=4$ state
if a line exists with a wavelength equal
to that of the $\mathcal{N}=4\rightarrow1$ transition (972.5~\AA).
In this case, the density required to
make the Br$\alpha$ line optically thick
would be lowered from the value $n\sim10^8$~cm$^{-3}$ given by
Equation~(\ref{eq:n4qn}).
Based on a detailed discussion
provided in Appendix~\ref{sec:ap4}, we found that
if a line with a wavelength of
$\sim$972.5~\AA\ and a transition probability
of $\sim10^{-1}$~s$^{-1}$
exists, we should take the resonant process into consideration 
in determining the population of the $\mathcal{N}=4$ state.

We searched for possible resonant lines with
wavelengths of $\sim$972.5~\AA\
within the velocity range $\sim10$~km~s$^{-1}$,
which corresponds to the thermal velocity at $T=10000$~K \citep{Osterbrock2006agn}
using the line list from the Cloudy code \citep{Ferland1998c9n}
and found no candidates for the X$_4$ line.
Thus, we conclude that the resonant process
does not take place
and is excluded from the excitation mechanisms
for the $\mathcal{N}=4$ state.

\subsubsection{Simulation with Cloudy}

In order to investigate quantitatively
the Br$\beta$/Br$\alpha$ line ratio
taking all excitation mechanisms into account,
we used the Cloudy code
\citep[ver.~10.00;][]{Ferland1998c9n}
and simulated the ratio
for the optically thick case.
Cloudy calculates the recombination
and the collisional processes altogether.
Cloudy also solves the radiative transfer of lines
and so can be used to investigate
the effect of optical depth
on the line fluxes.

To execute a simulation with Cloudy,
four parameters are required:
(1) the spectral shape of the incident radiation,
(2) the intensity of the incident radiation,
(3) the density of the surrounding gas,
and (4) the criterion for stopping the calculation.
We considered a single spherical \ion{H}{2} region
with a uniform gas ionized by a hot central star.
For the spectral shape of the incident radiation (1),
we used 40000~K blackbody radiation
to simulate a typical O star \citep{Osterbrock2006agn}.
For the intensity of the incident radiation (2),
we specified the number of ionizing photons per unit time, $Q(\mathrm{H})$.
We varied $Q(\mathrm{H})$ from $10^{48}$~s$^{-1}$
to $10^{51}$~s$^{-1}$ using intervals of a decade
on the assumption that the ionizing radiation
is dominated by massive OB stars in starburst regions.
The $Q(\mathrm{H})$ values $10^{48}$~s$^{-1}$,
$10^{49}$~s$^{-1}$, and $10^{50}$~s$^{-1}$
correspond to typical values for B stars,
O5 stars, and massive O stars (O3 stars),
respectively \citep{Osterbrock2006agn}.
We also calculated a case with $Q(\mathrm{H})=10^{51}$~s$^{-1}$
for reference.
We varied the gas density (3) from $n=10^5$~cm$^{-3}$
to $10^{10}$~cm$^{-3}$ using intervals of a decade.
To simulate line emission from the ionized region,
we adopted the electron fraction for the total gas, i.e., the degree of ionization,
to be 0.1 as the stopping criterion for the calculation (4).
In addition to the abovementioned parameters,
we specified a spherical geometry with the
inner radius of the surrounding gas,
which is required in Cloudy when we use $Q(\mathrm{H})$ as
the intensity of the incident radiation, equal to $r=10^{12}$~cm.
We iterated the calculations
until the difference between the line optical depths
of the last two iterations became smaller than 0.20.
The adopted parameters described above are summarized in Table~\ref{tab:clco}.
Other parameters are set to the default values of Cloudy; e.g.,
the line width is determined by the thermal velocity,
solar abundances are adopted,
and dust grains are not included in the calculations.

\begin{deluxetable*}{lrl}
\tablecaption{Parameters Used in Cloudy Simulations\tablenotemark{a} \label{tab:clco}}
\tablewidth{0pt}
\tablehead{
\colhead{Parameter}&\colhead{Value}&\colhead{Description}
}
\startdata
Blackbody&40000 K&Spectral shape of incident radiation.\\
log~$Q(\rm{H})$\tablenotemark{b}&48--51&Intensity of incident radiation.\\
log~$r$\tablenotemark{c}&12&Inner radius of surrounding gas.\\
log~$n$\tablenotemark{d}&5--10&Density of surrounding gas.\\
Stop Efrac\tablenotemark{e}&$-1$&Stopping criterion for calculation.\\
Sphere&\nodata&Geometry of surrounding gas.\\
Iterate to Convergence\tablenotemark{f}&0.20&Stopping criterion for iteration.
\enddata
\tablenotetext{a}{All input parameters we used for the calculations are listed.
Other parameters were set to the default values of Cloudy.}
\tablenotetext{b}{Log of the number of ionizing photons in s$^{-1}$.}
\tablenotetext{c}{Log of the inner radius of the gas in cm.}
\tablenotetext{d}{Log of the total hydrogen number density of the gas in cm$^{-3}$.}
\tablenotetext{e}{Log of the electron fraction below which the calculation stops.}
\tablenotetext{f}{The calculation is iterated until the difference between the optical
depths of the last two iterations becomes smaller than the specified value.}
\end{deluxetable*}

\begin{figure*}
\centering
\includegraphics[width=\textwidth]{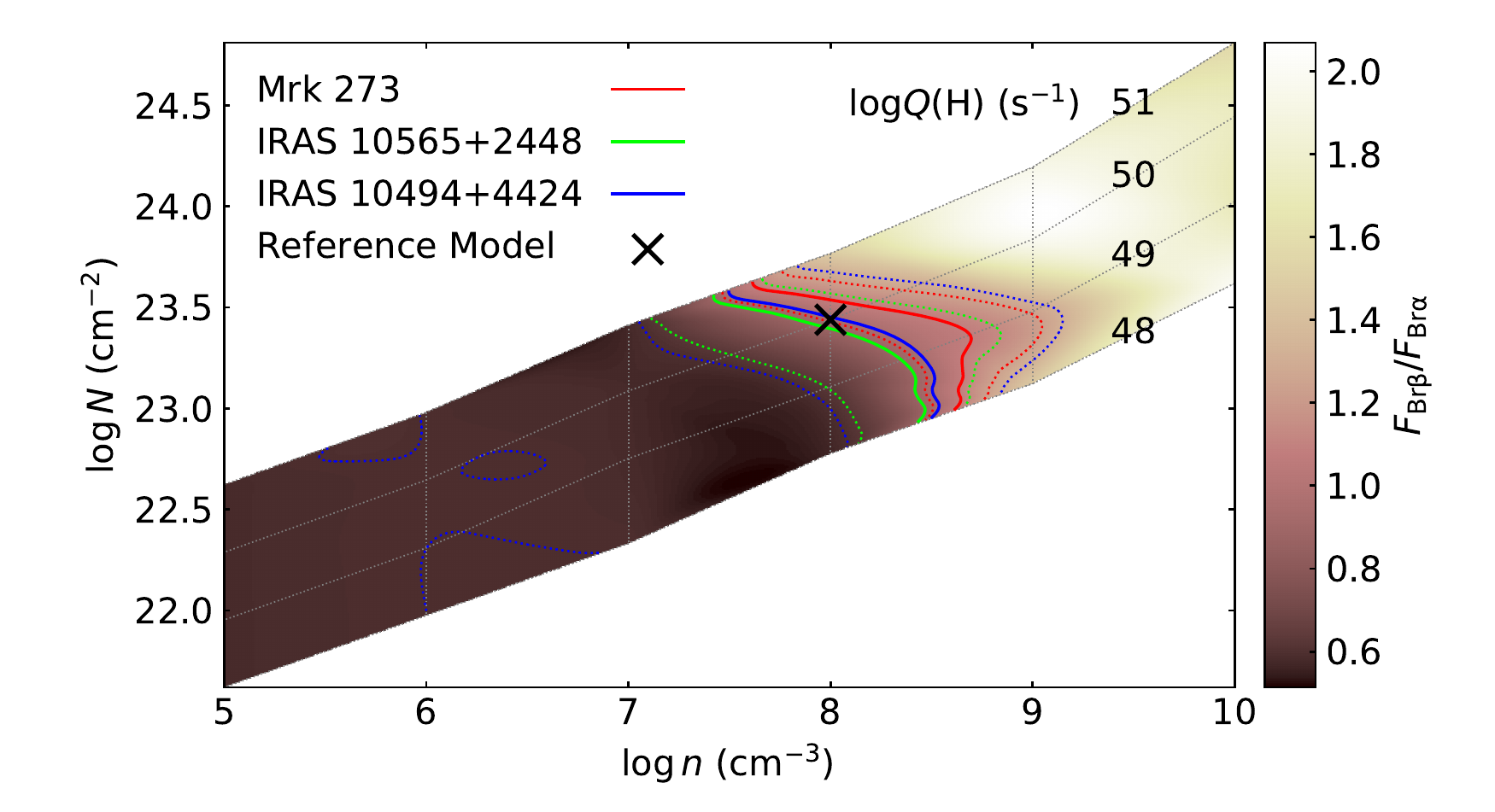}
\caption{Cloudy result for the Br$\beta$/Br$\alpha$ line ratios.
The number of ionizing photons $Q(\mathrm{H})$ and
the total hydrogen density $n$ are varied within the ranges
$10^{48}$~s$^{-1}\leq Q(\mathrm{H})\leq10^{51}$~s$^{-1}$
and $10^5$~cm$^{-3}\leq n\leq10^{10}$~cm$^{-3}$
using intervals of a decade.
The Br$\beta$/Br$\alpha$ line ratio is shown as the color scale.
The observed Br$\beta$/Br$\alpha$ line ratios for
Mrk~273, IRAS~10565$+$2448, and IRAS~10494$+$4424
are presented as red, green, and blue lines, respectively.
The observed values are shown by the solid lines,
and the ranges of 3$\sigma$ uncertainties are
indicated by the dotted lines.
The model calculated with $Q(\mathrm{H})=10^{50}$~s$^{-1}$
and $n=10^8$~cm$^{-3}$ is indicated by the black cross.
\label{fig:clres}}
\end{figure*}

The results of the Cloudy simulations
of the Br$\beta$/Br$\alpha$ line ratio
are shown in Figure~\ref{fig:clres}.
We find that the Br$\beta$/Br$\alpha$ line ratio
increases for high values of $n$ and $N$.
In contrast, the ratio
is close to the case B value (0.565)
for low values of $n$ and $N$.
The optical depth of the Br$\alpha$ line
is proportional to $N_4$, which is proportional
to $n$ and $N$, as shown in Equation~(\ref{eq:col4}).
Thus, the results are consistent with our simple estimate,
indicating that
a high Br$\beta$/Br$\alpha$ line ratio
is produced when the Br$\alpha$ line becomes
optically thick.

We now compare the observed Br$\beta$/Br$\alpha$ line ratios
with the Cloudy results.
Figure~\ref{fig:clres} indicates that
the observed ratios in the three galaxies
are consistent with $n\sim10^8$~cm$^{-3}$,
where the Br$\alpha$ line starts to become optically thick.
In conditions with higher values of $n$, the Br$\beta$/Br$\alpha$
line ratio becomes too large to match the observed anomalies.
From this result,
we conclude that in order to explain the observed Br$\beta$/Br$\alpha$ line ratio
within a single \ion{H}{2} region,
gas densities as high as $n\sim10^8$~cm$^{-3}$
are required to achieve column densities large enough
to make the Br$\alpha$ line optically thick.

\section{Comparison with Other Hydrogen Recombination Lines}
\label{sec:comp}

In Cloudy simulations, not only Br$\alpha$ and Br$\beta$,
but also other \ion{H}{1} recombination lines are calculated.
The intensity ratios between those lines can also be compared to observations.
We here review the existing observations of other \ion{H}{1} lines
in the optical and near-infrared
for the ULIRGs, in particular for the three objects
in which significant anomalous Br$\beta$/Br$\alpha$ ratios were found,
and demonstrate that they do not contradict the predictions of our high-density model.
This also explains why intensity ratio anomalies
such as the one revealed in this work
have not been found before.
In the following subsections,
we explain the reasons in detail for each \ion{H}{1} line of interest.
A basic short explanation is that
even if the intensity ratio of other \ion{H}{1} lines deviates from the case~B value
due to the high-density condition,
that change is indistinguishable from the effect of attenuation.

As a reference model for explaining
the observed Br$\beta$/Br$\alpha$ line ratio,
we adopt
the Cloudy result calculated
for $n=10^8$~cm$^{-3}$
and $Q=10^{50}$~s$^{-1}$.
This model represents
an \ion{H}{2} region, ionized by an O3 star
and surrounded by gas at high density.
The parameters and important results of this reference model
are summarised in Table~\ref{tab:rfm}.
The Br$\beta$/Br$\alpha$ line ratio
for this model (0.940) explains well
the observed values in 
IRAS~10494$+$4424, IRAS~10565$+$2448, and Mrk 273
of
$0.96\pm0.12$, $0.88\pm0.09$, and $1.09\pm0.05$, respectively.
The predictions made by this model
for several other \ion{H}{1} line ratios,
which are to be discussed below,
are listed in Table~\ref{tab:ratio}.
This table also shows the values for case~B
and the values observed in the three ULIRGs.

\begin{table}
\centering
\caption{Summary of the reference model\label{tab:rfm}}
\begin{tabular}{ll}
\hline\hline
Parameters & \\
\hline
Gas density        & $n=10^8~\mathrm{cm}^{-3}$ \\
Incident radiation & $Q(\mathrm{H})=10^{55}~\mathrm{s}^{-1}$ \\
\hline
Results & \\
\hline
Column density        & $N=2.75\times10^{23}~\mathrm{cm}^{-2}$ \\
Br$\alpha$ luminosity & $L_{\mathrm{Br}\alpha}=2.79\times10^{36}~\mathrm{erg~s^{-1}}$ \\
Br$\beta$/Br$\alpha$ ratio & $F_\mathrm{Br\beta}/F_\mathrm{Br\alpha}=0.940$ \\
\hline
Case B value & $F_\mathrm{Br\beta}/F_\mathrm{Br\alpha}=0.565$ (for reference)\\
\hline
\end{tabular}
\end{table}

\begin{deluxetable*}{cccccc}
\tablecaption{Modeled and Observed \ion{H}{1} Line Ratios \label{tab:ratio}}
\tablewidth{0pt}
\tabletypesize{\small}
\tablehead{
\colhead{Lines}&\colhead{Case B}&\colhead{High-Density}
&\colhead{IRAS}&\colhead{IRAS}&\colhead{Mrk}\\
\colhead{}&\colhead{}&\colhead{Model}&
\colhead{10494$+$4424}
&\colhead{10565$+$2448}&\colhead{273}
}
\startdata
Br$\beta$/Br$\alpha$  & 0.565 & 0.940 & $0.96\pm0.12$                 & $0.88\pm0.09$   & $1.09\pm0.05$                 \\
H$\alpha$/Br$\alpha$  & 34.4  & 37.6  & $1.34\pm0.17$                 & $2.39\pm0.28$   & $6.70\pm0.69$                 \\
H$\beta$/H$\alpha$    & 0.348 & 0.216 & $0.110\pm0.005$               & $0.061\pm0.003$ & $0.100\pm0.005$               \\
Pa$\alpha$/Br$\gamma$ & 12.1  & 9.93  & $16.1\pm1.5$\tablenotemark{a} & \nodata         & $10.8\pm1.0$\tablenotemark{b} \\
\enddata
\tablenotetext{a}{Taken from \cite{Murphy2001kbs}.}
\tablenotetext{b}{Taken from \cite{Veilleux1999n}.}
\end{deluxetable*}

\subsection{Comparison with H$\alpha$ Line}

In this subsection, we
compare the fluxes of the Brackett lines and the H$\alpha$ line
for IRAS~10494$+$4424, IRAS~10565$+$2448, and Mrk~273.
For IRAS~10494$+$4424 and Mrk~273,
we use the H$\alpha$ line flux obtained from our
Nickel observations (Table~\ref{tab:haflux}).
IRAS~10565$+$2448 was observed
with the integral field unit on the Gemini North telescope by \cite{Shih2010c};
they reported an integrated H$\alpha$ line flux
$F_{\mathrm{H}\alpha}=(8.42\pm0.84)\times10^{-14}$~erg~s$^{-1}$~cm$^{-2}$
for this galaxy
(before extinction correction) within an aperture of $5''\times7''$.
We adopt this flux for comparison with our results for IRAS~10565$+$2448.

\begin{figure}
\plotone{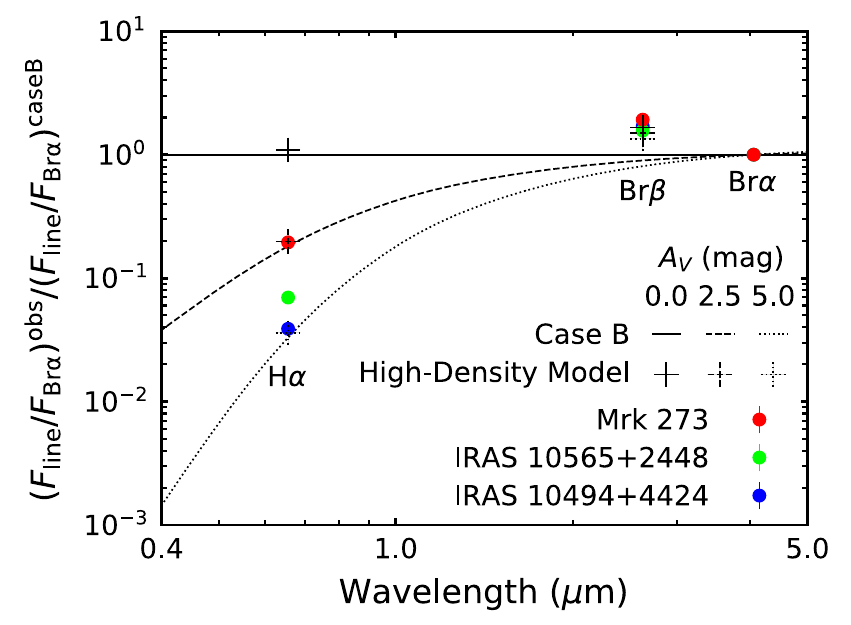}
\caption{Hydrogen recombination-line fluxes
relative to the Br$\alpha$ line.
The ratios are normalized to those
predicted for case~B
\citep[10000~K, low-density limit:][]{Osterbrock2006agn}.
The black lines show the line ratios
for the case~B model with different visual extinctions.
The black crosses show the line ratios predicted by the high-density model.
The solid, dashed, and dotted crosses indicate
the ratios with no dust extinction, with extinction of $A_V=2.5$~mag, and
with extinction of $A_V=5.0$~mag, respectively.
\label{fig:hline}}
\end{figure}

We summarize the H$\alpha$/Br$\alpha$ line ratio in Table~\ref{tab:ratio}.
Figure~\ref{fig:hline} compares the flux of
the Br$\alpha$, Br$\beta$, and H$\alpha$ lines.
The high-density model
explains well both the H$\alpha$/Br$\alpha$ and
Br$\beta$/Br$\alpha$ line ratios if the dust extinctionis $A_V\sim2.5$--$5.0$~mag;
this is higher than the typical value
of $A_V\sim1.3$~mag suggested by \cite{Kennicutt1998sfi}
for typical spiral galaxies.
The Br$\beta$/Br$\alpha$ line ratio
is not much affected (reduced only by $\sim10$--20\%),
even for relatively high dust extinction.
Thus, the high Br$\beta$/Br$\alpha$ line ratio predicted by the high-density model
is clearly seen even with dust extinction of $A_V\sim2.5$--$5.0$~mag,
which explains well the observed ratios.
In contrast,
the H$\alpha$/Br$\alpha$ line ratio
is determined almost exclusively by the effect of dust extinction,
and the difference between case B and the high-density model is small.
The observed H$\alpha$/Br$\alpha$ line ratio
is consistent with the high-density model if the dust extinction is
$A_V\sim2.5$--$5.0$~mag,
and it is also explained by the case~B model with almost the same dust extinction.
We cannot distinguish between these two models from the H$\alpha$/Br$\alpha$ line ratio.

We conclude that the high-density model is
consistent with the observations of the H$\alpha$ line.
The effect of dust
extinction at optical wavelengths
is so strong that the deviation of the \ion{H}{1} line ratios
from those of case~B
due to the high-density condition
is easily cancelled out and made unnoticeable.

\subsection{Comparison with H$\beta$/H$\alpha$ Line Ratio}
\label{sec:cwh}

The H$\beta$ line ($\mathcal{N}=4\rightarrow2$, 4861~\AA)
is also observable in the optical as well as H$\alpha$,
and the H$\beta$/H$\alpha$ line ratio is one of the most
intensively studied ratios among \ion{H}{1} lines \citep[e.g.,][]{Kim1998i1j}.
We here compare the H$\beta$/H$\alpha$ line ratio with our
Br$\beta$/Br$\alpha$ line ratio.

For the AGNUL sample,
we use the H$\beta$/H$\alpha$ line ratio
summarized in Table~3 of \cite{Yano2016s}.
The ratios for the two galaxies added to our
sample in this paper are summarized
in Appendix~\ref{sec:nuliz}.
The H$\beta$/H$\alpha$ line ratios for all the targets
are well below the case~B value of 0.348 (Table~\ref{tab:ratio}).
This indicates that the H$\beta$/H$\alpha$ line ratio
can be explained with a combination of case~B and dust extinction.
The inferred dust extinction in this case
($A_V^\mathrm{opt}$ in the tables)
is typically $\sim2.5$~mag.

\begin{figure}
\plotone{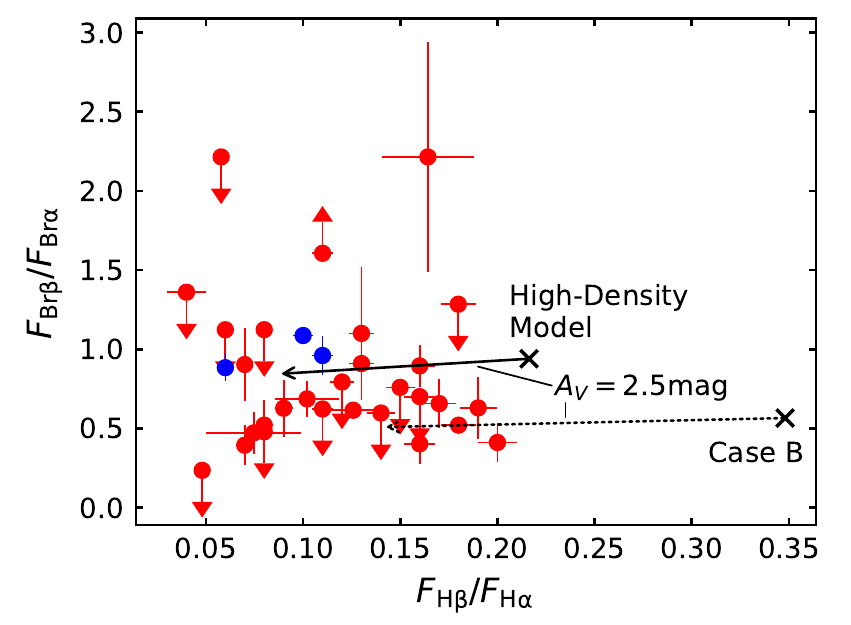}
\caption{Br$\beta$/Br$\alpha$ line ratio versus H$\beta$/H$\alpha$ line ratio.
The three galaxies with high Br$\beta$/Br$\alpha$ line ratios
are represented by blue circles, while the others are shown as red circles.
The black cross indicates the line ratios predicted by the high-density model.
The extinction vectors for $A_V=2.5$ mag are shown as the black arrows.
The solid arrow shows extinction from the ratios of the high-density model,
whereas the dashed arrow indicates extinction from the ratios of case~B.
\label{fig:brvsopt}}
\end{figure}

A comparison of the H$\beta$/H$\alpha$ line ratio
with our Br$\beta$/Br$\alpha$ line ratio is shown in Figure~\ref{fig:brvsopt}
and displays a large scatter between the two ratios.
The extinction vectors in Figure~\ref{fig:brvsopt} indicate that
the H$\beta$/H$\alpha$ line ratio is strongly affected by dust extinction.
In contrast, the Br$\beta$/Br$\alpha$ line ratio
is almost unchanged with dust extinction of $A_V\sim2.5$~mag.

The observed high Br$\beta$/Br$\alpha$ line ratios
cannot be achieved by the case~B model.
In contrast, the extinction vector
from the high-density model indicates that
both the high Br$\beta$/Br$\alpha$ line ratio
and the H$\beta$/H$\alpha$ line ratio
are well explained by the high-density model
with dust extinction of $A_V\sim2.5$~mag.

In the high-density model,
the H$\beta$/H$\alpha$ line ratio
is predicted to be 0.216 (Table~\ref{tab:ratio}).
We note that the deviation of
the H$\beta$/H$\alpha$ line ratio
in the high-density model is
not distinguishable from
the effect of dust extinction.
This indicates that the deviation
cannot be probed with
the H$\beta$/H$\alpha$ line ratio
even in conditions wherein dust extinction is very low.

From these results,
we conclude that the high-density model
is consistent with the observations of the H$\beta$/H$\alpha$ line ratio.
We also conclude that the deviation of the \ion{H}{1} line ratios
from those of case~B
can be observed only in the infrared,
wherein the effect of dust extinction is small.

\subsection{Comparison with Pa$\alpha$/Br$\gamma$ Line Ratio}

In this subsection, we compare our results with infrared \ion{H}{1} lines.
In the infrared,
the Br$\gamma$ line at 2.17~$\mu$m
and the Pa$\alpha$ line at 1.88~$\mu$m
are both observable in $K$-band ($\sim1.9$--$2.5$~$\mu$m)
observations from the ground
for nearby objects ($z\sim0.01$--$0.15$).
These lines are relatively strong
and have been observed
in various studies to measure dust extinction
\citep[e.g.,][]{Goldader1995sli,Veilleux1999n}.

The Pa$\alpha$/Br$\gamma$ line ratio predicted by the high-density model is 9.93,
deviating from the case B value of 12.1 (Table~\ref{tab:ratio}).
In the prediction,
the Pa$\alpha$ line at the shorter wavelength
is weakened relative to the Br$\gamma$ line
at the longer wavelength.
This is the same trend as the effect of dust extinction.
The Pa$\alpha$/Br$\gamma$ line ratio of 9.93
corresponds to dust extinction of $A_V=6.60$~mag,
assuming the case B ratio.
This indicates that we cannot
distinguish the deviation of
the predicted Pa$\alpha$/Br$\gamma$ line ratio from that of case B
due to the effect of dust extinction.
Thus, our model is consistent with
the fact that no anomaly has been reported
in previous observations
of the Pa$\alpha$/Br$\gamma$ line ratio.

Among our targets with
anomalous Br$\beta$/Br$\alpha$ line ratios,
the Pa$\alpha$/Br$\gamma$ line ratio
was observed in IRAS~10494$+$4424
by \cite{Murphy2001kbs} and 
in Mrk 273 by \cite{Veilleux1999n}.
We cannot find observations
of the Pa$\alpha$/Br$\gamma$ line ratio
in IRAS~10565$+$2448.
The Pa$\alpha$/Br$\gamma$ line ratio in
Mrk 273 is $10.8\pm1.0$, which
is consistent with our prediction.
The Pa$\alpha$/Br$\gamma$ line ratio in
IRAS~10494$+$4424 was reported to be $16.1\pm1.5$.
This ratio indicates that the Pa$\alpha$ line
is enhanced relative to the Br$\gamma$ line
compared to the case B ratio.
This is opposite to the effect of dust extinction
and also to the deviation of our model from the case B ratio.
Thus, the ratio is explained neither
by our model
nor by the case B ratio with dust extinction.
\cite{Murphy2001kbs}
regarded this anomaly as not significant.
They claimed that the observations
of the Pa$\alpha$ and Br$\gamma$ lines
in IRAS~10494$+$4424 were performed with
different apertures, and this made
the observed Pa$\alpha$/Br$\gamma$ line ratio
uncertain by as much as 50\%.
Thus, we also treat the deviation as not significant here.

Based on the above results,
we conclude that our model 
does not contradict previous observations of
the \ion{H}{1} line ratios.
The deviation of a line intensity ratio from case B due to the high density condition is
hardly detectable in ratios other than Br$\beta$/Br$\alpha$,
in the optical and near-infrared wavelength regions.

\section{Structure of High-Density \ion{H}{2} Regions}
\label{sec:str}

We find that we can explain the high Br$\beta$/Br$\alpha$
line ratio with the optical-depth effect.
In order to make the Br$\alpha$ line optically thick,
high-density \ion{H}{2} regions are required in our Cloudy model.
Herein, we discuss possible structures of such
high-density \ion{H}{2} regions in ULIRGs.

\subsection{Two Extreme Cases}

To explain the observed luminosity of the Br$\alpha$ line,
we consider two extreme cases for the line-emitting regions:
(1) an ensemble of \ion{H}{2} regions, each ionized by a single star
and (2) a giant \ion{H}{2} region where all the ionizing stars are
concentrated at the center.
For each case, we discuss whether
the optical-depth effect can
produce high Br$\beta$/Br$\alpha$ line ratios.

\subsubsection{Ensemble of \ion{H}{2} Regions}
\label{sec:ens}

We here consider an ensemble of \ion{H}{2} regions,
each of which is represented by the high-density model.
In Table~\ref{tab:size}, we list the observed luminosity of
the Br$\alpha$ line for each of the three galaxies
with a high Br$\beta$/Br$\alpha$ line ratio.
In the high-density model,
the luminosity of the Br$\alpha$ line
produced by a single \ion{H}{2} region is found to be
$L_{\mathrm{Br}\alpha}^\mathrm{Model}=2.79\times10^{36}$~erg~s$^{-1}$ (Table~\ref{tab:rfm}),
and we estimate the number of \ion{H}{2} regions $k_\mathrm{tot}$,
as shown in Table~\ref{tab:size}.
From this result, we conclude that $\sim10^5$ \ion{H}{2} regions
with high-density conditions
are required to explain our observations.

\begin{deluxetable}{ccc}
\tablecaption{Comparison of Luminosity with High-Density Model \label{tab:size}}
\tablewidth{0pt}
\tablehead{
\colhead{Object}&\colhead{$L_{\mathrm{Br}\alpha}$}&\colhead{$k_\mathrm{tot}$\tablenotemark{a}}\\
\colhead{Name}&\colhead{($10^{41}\;$erg\ s$^{-1}$)}&\colhead{($10^4$)}
}
\startdata
IRAS~10494$+$4424 & $2.31\pm0.18$ & 8.8 \\
IRAS~10565$+$2448 & $1.50\pm0.09$ & 5.7 \\
Mrk~273           & $1.58\pm0.04$ & 6.0 \\
\enddata
\tablenotetext{a}{The number of \ion{H}{2} regions represented by the high-density model
that are required to yield the observed luminosity of the Br$\alpha$ line.}
\end{deluxetable}

We next estimate the expected number of \ion{H}{2} regions
along the line of sight, $k_\mathrm{los}$, to determine the effect of optical depth
on the line emitted from a single \ion{H}{2} region
and intercepted by other \ion{H}{2} regions.
We assume a volume filling factor $\varepsilon\sim10^{-6}$,
which is a typical value for \ion{H}{2} regions
observed in starburst galaxies \citep{Anantharamaiah1993d}.
We also assume that \ion{H}{2} regions are
uniformly and spherically distributed.
Using the radius $R$ of individual \ion{H}{2} regions,
the volume of the entire space in which they are distributed as an ensemble
can be written as $k_\mathrm{tot}R^3/\varepsilon$.
Then, the diameter $d$ of that space becomes
\begin{equation}
d\sim(k_\mathrm{tot}/\varepsilon)^{1/3}R. \label{eq:dktot}
\end{equation}
On the other hand,
$k_\mathrm{los}$, the number of \ion{H}{2} regions that exist on the diameter, satisfies
\begin{equation}
k_\mathrm{los}R^3 \sim \varepsilon d R^2. \label{eq:dklos}
\end{equation}
By comparing Equations~(\ref{eq:dktot}) and (\ref{eq:dklos}), we find
\begin{equation}
k_\mathrm{los}\sim{k_\mathrm{tot}}^{\frac{1}{3}}\varepsilon^{\frac{2}{3}}. \label{clos}
\end{equation}
Substituting $k_\mathrm{tot}=10^5$ and $\varepsilon=10^{-6}$ into Equation~(\ref{clos}),
we obtain $k_\mathrm{los}\sim10^{-7/3}$.
We assume the relative velocities of \ion{H}{2} regions
to be of the order of $\sim100$~km~s$^{-1}$,
which is a typical line velocity observed in galaxies \citep[e.g.,][]{Osterbrock2006agn}
and is an order of magnitude higher than the thermal velocity.
Then, the optical depth of the Br$\alpha$ line caused by the intercepting \ion{H}{2} regions
is found to be three orders of magnitude smaller than that
of the \ion{H}{2} region from which the line originates.
Thus, we conclude that the line ratio produced in a single \ion{H}{2} region
is not affected by other \ion{H}{2} regions,
even if we consider an ensemble of $\sim10^5$ \ion{H}{2} regions.

We therefore conclude that an ensemble of \ion{H}{2} regions,
in each of which the Br$\alpha$ line is optically thick,
can explain the high Br$\beta$/Br$\alpha$ line ratio.
This ratio
is produced within each \ion{H}{2} region,
and what we observe is a collection of such \ion{H}{2} regions.
To achieve a column density large enough
to make the Br$\alpha$ line optically thick within a single \ion{H}{2} region,
the gas density must be as high as $n\sim10^8$~cm$^{-3}$.

\subsubsection{Single Giant \ion{H}{2} Region}
\label{sec:sgh}

We next consider another simplified model,
in which the line-emitting region is not a collection of \ion{H}{2} regions
but a single giant \ion{H}{2} region,
where all the ionizing stars are concentrated at the center of a uniform gas.
We investigate whether a high Br$\beta$/Br$\alpha$ line ratio
can be produced within such a giant \ion{H}{2} region by the optical-depth effect.

We have shown that ionizing sources with $Q(\mathrm{H})$ of the order of $\sim10^{55}$~s$^{-1}$
are required to explain the observed luminosity of the Br$\alpha$ line.
Thus, we also assume a central ionizing source of $Q(\mathrm{H})=10^{55}$~s$^{-1}$,
which corresponds to 10$^5$--10$^6$ OB stars.
For the turbulence velocity within the giant \ion{H}{2} region,
we assume $v_\mathrm{Turb}=100$~km~s$^{-1}$.
In this case, Equation~(\ref{eq:taua}) indicates that
$N_4\sim2\times10^{12}$~cm$^{-2}$ is required
to make the Br$\alpha$ line optically thick.
Substituting $N_4=2\times10^{12}$~cm$^{-2}$
and $Q(\mathrm{H})=10^{55}$~s$^{-1}$ into Equation~(\ref{eq:n4qn}),
we obtain $n\sim2\times10^7$~cm$^{-3}$ for the gas density
required to make the Br$\alpha$ line optically thick
within the giant \ion{H}{2} region.

\begin{figure}
\plotone{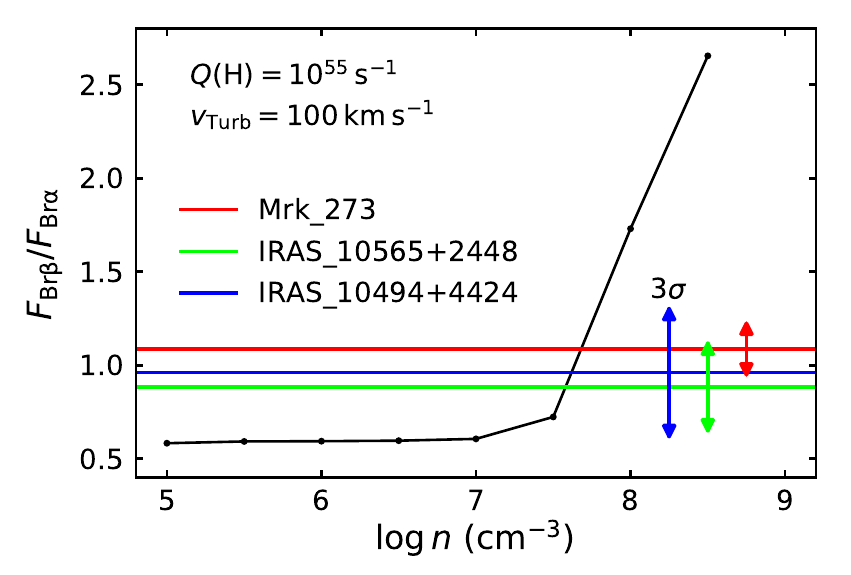}
\caption{Cloudy results for the Br$\beta$/Br$\alpha$ line ratios
for various total hydrogen densities $n$,
with $Q(\mathrm{H})=10^{55}$~s$^{-1}$ and $v_\mathrm{Turb}=100$~km~s$^{-1}$.
The horizontal red, green, and  blue lines show the observed $F_{\mathrm{Br}\beta}/F_{\mathrm{Br}\alpha}$
ratios for Mrk~273, IRAS~10565$+$2448, and IRAS~10494$+$4424, respectively.
The range of 3$\sigma$ uncertainty is shown by the vertical arrows.
\label{fig:tur}}
\end{figure}

To investigate the Br$\beta$/Br$\alpha$ line ratio quantitatively for this case,
we again used the Cloudy code to simulate the giant \ion{H}{2} region.
Most of the parameters are the same as
those tabulated in Table~\ref{tab:clco},
except that $Q(\mathrm{H})$ is now fixed at $10^{55}$~s$^{-1}$,
$n$ is varied at intervals of 0.5 dex up to $n=10^{8.5}~\mathrm{cm}^{-2}$,
and $v_\mathrm{Turb}=100$~km~s$^{-1}$ is adopted.
We show the Cloudy results in Figure~\ref{fig:tur}.
The observed Br$\beta$/Br$\alpha$ line ratio
can be explained if the gas density is
within the range $n=10^{7.5}$--$10^{8.0}$~cm$^{-3}$.
The luminosity of the Br$\alpha$ line
in these conditions is $L_{\mathrm{Br}\alpha}\sim$2--4$\times10^{41}$~erg~s$^{-1}$,
and we confirm that this agrees well with the observed value.
We thus conclude that gas densities as high as $n\sim10^8$~cm$^{-3}$
are also required to explain the observed Br$\beta$/Br$\alpha$ line ratio
in the extreme case in which the line-emitting region is a single
giant \ion{H}{2} region with a central ionizing source emitting $Q(\mathrm{H})=10^{55}$~s$^{-1}$.

There is one caveat to the Cloudy results shown in Figure~\ref{fig:tur}.
With $Q(\mathrm{H})=10^{55}$~s$^{-1}$,
the column density of electrons within the \ion{H}{2} region
exceeds $N_e\sim2\times10^{24}$~cm$^{-2}$
for conditions with $n\geq10^8$~cm$^{-3}$
so the \ion{H}{2} region becomes Compton thick.
The Cloudy code is not designed to simulate Compton-thick regimes \citep{Ferland1998c9n};
therefore, the validity of the result is not guaranteed in those conditions.
The process of Thomson scattering does not include any energy transfer.
Thus, we expect that the result for the emergent line ratio is still valid
even if the \ion{H}{2} region becomes Compton thick.

In summary,
for both the two extreme cases,
we conclude that gas densities as high as $n\sim10^8$~cm$^{-3}$
are required to achieve a column density of neutral hydrogen large enough
to make the Br$\alpha$ line optically thick.
We propose this high-density scenario as the most plausible cause
of the high Br$\beta$/Br$\alpha$ line ratio.

\subsection{Impact of dust grains}

The density of $n\sim10^8~\mathrm{cm}^{-3}$ is in a regime
denser than the densest ultracompact \ion{H}{2} regions in our Galaxy
\citep[e.g.,][]{Kurtz2000u,Churchwell2002u}.
In this regime, dust can compete with \ion{H}{1} for ionization photons.
Then, the effective $Q(\mathrm{H})$ for \ion{H}{1} is reduced,
and a higher number density may be needed
to achieve a column density sufficient to make the Br$\alpha$ line optically thick
(Equation (\ref{eq:n4qn})).
On the other hand, it is also expected that the higher the density,
the more the dust is thermally coupled with gas,
and the abundance of the dust decreases due to sublimation.
The overall influence of the dust includes both of these effects.
Because the above Cloudy models are assumed to be dust-free,
we here quantitatively evaluate how the results change when the dust is taken into account
by modifying the cloudy models.

\begin{figure}
\plotone{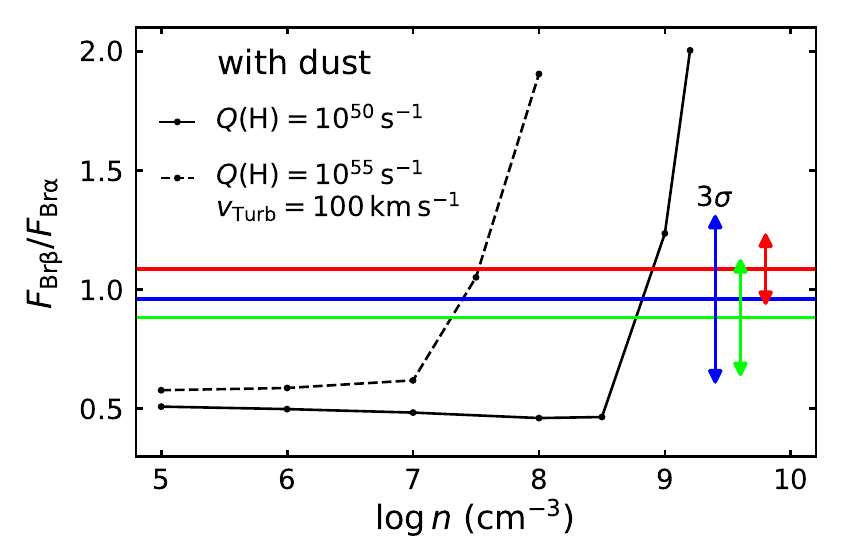}
\caption{Cloudy results for the Br$\beta$/Br$\alpha$ line ratio
obtained when dust grains are included in the simulation.
The ratio is calculated as a function of total hydrogen densities $n$.
The black solid line shows the result at $Q(\mathrm{H})=10^{50}$~s$^{-1}$,
which corresponds to the case of an ensemble of \ion{H}{2} regions (\S\ref{sec:ens}).
The black dashed line shows the result at $Q(\mathrm{H})=10^{55}$~s$^{-1}$
with $v_\mathrm{Turb}=100$~km~s$^{-1}$,
which corresponds to the case of a single giant \ion{H}{2} region (\S\ref{sec:sgh}).
The horizontal red, green, and  blue lines show the observed $F_{\mathrm{Br}\beta}/F_{\mathrm{Br}\alpha}$
ratios for Mrk~273, IRAS~10565$+$2448, and IRAS~10494$+$4424, respectively.
The range of 3$\sigma$ uncertainty is shown by the vertical arrows.
\label{fig:cldust}}
\end{figure}

We start with the case of an ensemble of \ion{H}{2} regions.
Cloudy simulations were performed with dust.
The abundances were changed from the default to a predefined set of the interstellar medium.
Both silicate and graphite grains are included with ten size bins for each.
The option to treat the dust sublimation was turned on
so that the abundance of each grain species steeply decreases
when its temperature is above the sublimation temperature. 
The Br$\beta$/Br$\alpha$ line ratio was calculated as a function of $n$.
To examine the case of the intensest ionization, $Q(\mathrm{H})$ was
fixed at $10^{50}~\mathrm{s}^{-1}$ (corresponding to an O3 star).
Other parameters are the same as in Table~\ref{tab:clco}.
The results are shown in Figure~\ref{fig:cldust}.
With the same parameters as the reference model in the dust-free case
($n=10^8~\mathrm{cm}^{-3}$, $Q(\mathrm{H})=10^{50}~\mathrm{s}^{-1}$),
the Br$\beta$/Br$\alpha$ ratio is lowered from 0.94 to 0.46.
This is due to the fact that the column density $N$ is not much increased
and Br$\alpha$ does not become optically thick
because ionizing photons are consumed by dust,
and that the ratio is affected by dust extinction.
The ratio of $\sim$1 as observed is found at $n=10^9~\mathrm{cm}^{-3}$.
At this point, $L_{\mathrm{Br}\alpha}^\mathrm{Model}$ becomes
$2.4\times10^{35}$~erg~s$^{-1}$,
which is a factor of 12 lower than in the dust-free reference model.
Thus, the estimate of $k_\mathrm{tot}$ increases up to $10^6$.
The estimate of $k_\mathrm{los}$ changes by only a factor of two,
and the possibility that multiple \ion{H}{2} regions are aligned along the line of sight
and the line emission from the \ion{H}{2} region behind is blocked is still negligible.

Next, we move on to the case of a single giant \ion{H}{2} region.
In the above case of the \ion{H}{2} region ensemble with dust,
$k_\mathrm{tot}$ has increased by one order of magnitude.
However, in this case, the $Q(\mathrm{H})$ needed is expected to be
the same as in the dust-free model, $Q(\mathrm{H})=10^{55}~\mathrm{s}^{-1}$.
This is because if multiple stars are concentrated near the center,
they will destroy dust grains in the same region together,
and thus cancel out the dust effect more efficiently than in the ensemble case,
where each star destroys the surrounding dust on its own. 
We thus changed the $Q(\mathrm{H})$ of the above Cloudy model with dust
to $10^{55}~\mathrm{s}^{-1}$
and calculated the Br$\beta$/Br$\alpha$ ratio as a function of $n$.
The specification of $v_\mathrm{Turb}=100$~km~s$^{-1}$ was also added.
The results are shown in Figure~\ref{fig:cldust}.
A ratio of $\sim$1 is found at $n=10^{7.5}~\mathrm{cm}^{-3}$.
Therefore, a high-density situation is still necessary.
At this point, $L_{\mathrm{Br}\alpha}^\mathrm{Model}=1\times10^{42}$~erg~s$^{-1}$,
which fully explains the observed line luminosities (Table~\ref{tab:size}).
Hence, as predicted, in the case of a giant \ion{H}{2} region,
the observation results can be explained by
$Q(\mathrm{H})=10^{55}~\mathrm{s}^{-1}$ ($\sim10^5$ O3 stars)
even if dust is taken into consideration.

The actual situation is expected to be bracketed between these two extreme cases.
Therefore, we conclude that the number of massive stars
needed to explain the Br$\beta$/Br$\alpha$ anomaly may be increased
by up to an order of magnitude from the estimates
based on the assumption of the dust-free gas.
The main conclusion
(the high density gas is needed to explain the anomalous \ion{H}{1} line ratio)
is the same for the cases even with dust.

\subsection{Comparison of Lifetime with Galactic Ultracompact \ion{H}{2} Regions}

In the dust-free ensemble model, about
$10^5$ \ion{H}{2} regions
with gas at high densities $n\sim10^8$~cm$^{-3}$
are required
to explain the high Br$\beta$/Br$\alpha$ line ratios.
Herein, we discuss the possibility of
observing so many high-density \ion{H}{2} regions
in ULIRGs
by comparison with the ultracompact \ion{H}{2} regions in our Galaxy.

Ultracompact \ion{H}{2} regions in our Galaxy
\citep[e.g.,][]{Kurtz2000u,Churchwell2002u}
are reported to contain high-density gas
up to $n\sim10^7$~cm$^{-3}$
and to have sizes of the order of $10^{-2}$~pc
\citep[e.g.,][]{dePree1995u}.
The densities and sizes of
these ultracompact \ion{H}{2} regions are
comparable to those of the high-density \ion{H}{2} regions
required in our model.
We thus consider the high-density \ion{H}{2} regions in our model
as analogous to the ultracompact \ion{H}{2} regions in our Galaxy.

A simple estimate of the lifetime of an \ion{H}{2} region in an ultracompact
state with a size $r\sim10^{-2}$~pc
can be obtained by dividing $r$ by the sound speed $v_\mathrm{s}$
in the ionized material ($v_\mathrm{s}\sim10$~km~s$^{-1}$ at $T=10000$~K),
assuming that the \ion{H}{2} region
expands at a speed comparable to $v_\mathrm{s}$.
This yields a lifetime $r/v_\mathrm{s}\sim10^{3}$~yr
for an ultracompact \ion{H}{2} region in our Galaxy.
Another estimate can be obtained from
the number of ultracompact \ion{H}{2} regions,
which is estimated to be $\sim10^3$ in our Galaxy \citep{Churchwell2002u}.
Adopting the formation rate of O stars
to be $\sim10^{-2}$~stars~yr$^{-1}$ in our Galaxy \citep{dePree1995u},
we get instead for the lifetime of an ultracompact \ion{H}{2} region $\sim10^{5}$,
some two orders of magnitude longer
than that obtained from
the simple expansion of an \ion{H}{2} region.
This large difference between the estimated lifetimes is
recognized as the ``lifetime problem,''
first mentioned by \cite{Wood1989m},
which still remains an open question \citep[e.g.,][]{Kurtz2000u,Churchwell2002u}.
Herein, we simply adopt $\sim10^5$~yr
as the representative lifetime of an ultracompact \ion{H}{2} region.

We assume that the high-density \ion{H}{2} regions
required in our high-density model
have a lifetime of the same order ($\sim10^5$~yr)
as those of the ultracompact \ion{H}{2} regions in our Galaxy.
We note that the star formation rate (SFR) in a ULIRG is
about two orders of magnitude higher than
that in our Galaxy \citep{Sanders1988uig}.
Scaling the number of the ultracompact
\ion{H}{2} regions observed in our Galaxy
\citep[$\sim10^3$;][]{Churchwell2002u}
with the SFR,
we thus expect the number of high-density
\ion{H}{2} regions to be on the order of $\sim10^5$ in a ULIRG.
Thus, we conclude that it is indeed possible to
have $\sim10^5$ high-density \ion{H}{2} regions
in a ULIRG, as our model predicts.

After the $\sim10^5$~yr lifetime of the ultracompact phase,
the gas density of an \ion{H}{2} region is expected to
fall below $\sim10^7$~cm$^{-3}$
as the \ion{H}{2} region expands \citep{dePree1995u}.
The typical lifetime of an O star is on the
order of $10^6$~yr \citep{Osterbrock2006agn}, which is an order of magnitude
longer than that of the ultracompact \ion{H}{2} regions.
Thus, the number of \ion{H}{2} regions
with gas densities lower than $\sim10^7$~cm$^{-3}$
is expected to be an order of magnitude larger than
that of the ultracompact \ion{H}{2} regions.
However, to explain the high Br$\beta$/Br$\alpha$ line ratios,
our model requires most \ion{H}{2} regions
to be in the ultracompact phase.
Our results thus indicate that some mechanism is required to
ensure that the Brackett lines 
from the \ion{H}{2} regions in ULIRGs are dominated by those emitted from
ultracompact \ion{H}{2} regions.
This problem remains when the dust is taken into account,
where a larger number of denser \ion{H}{2} regions are required
if they are isolated from each other.

\section{Prediction to Radio Recombination Lines}
\label{sec:imp}

We consider the effect of high densities
on radio recombination lines.
Our model requires gas densities $n=10^8$~cm$^{-3}$.
At such high densities,
collisional processes become important
for high-$\mathcal{N}$ states.
As shown in Figure~\ref{fig:depcoef},
hydrogen levels with low principal
quantum numbers ($\mathcal{N}\leq15$)
are not dominated by collisions,
even at a density $n=10^8$~cm$^{-3}$.
In contrast, the collisional processes
start to contribute significantly for states with
$\mathcal{N}\geq20$.
This indicates that hydrogen radio recombination
lines emitted with transitions
involving high-$\mathcal{N}$ states
are affected by the high densities our model predicts.

\cite{Peters2012u} showed that at
a density $n=10^8$~cm$^{-3}$,
$b_\mathcal{N}\sim 1$ for levels with
$\mathcal{N}>30$.
For $n=10^6$~cm$^{-3}$,
where anomalous
Br$\beta$/Br$\alpha$ line ratios are not found
in the Cloudy simulations,
only states with $\mathcal{N}>50$
become thermalized.
This indicates that observations of
radio recombination lines
with low-$\mathcal{N}$ transitions
($\mathcal{N}<50$) are required
to probe our model predictions.
However, previous observations
of radio recombination lines
were mainly focused on high-$\mathcal{N}$
transitions because of difficulty
of observations at the high frequencies
where the transitions with $\mathcal{N}<50$ are located.

Radio recombination lines
involving levels with $\mathcal{N}<50$
are now observable
within the frequency range of ALMA. 
We thus predict that such radio recombination lines
will be found to be thermalized
in those galaxies with Br$\beta$/Br$\alpha$ line-ratio anomalies, 
although the radio recombination lines are generally very weak
\citep[e.g.,][]{Izumi2016, Michiyama2020sft}. 

\section{Summary}
\label{sec:sum}

We conducted systematic observations
of the \ion{H}{1} Br$\alpha$ and Br$\beta$ lines
with the \textit{AKARI} IRC
for 52 nearby ($z<0.3$) ULIRGs.
We detected Br$\alpha$ and Br$\beta$ lines in
33 ULIRGs.
Among these,
three galaxies, IRAS~10494$+$4424,
IRAS~10565$+$2448, and Mrk~273,
show Br$\beta$/Br$\alpha$
line ratios ($0.96\pm0.12$, $0.883\pm0.085$, and $1.086\pm0.053$, respectively),
which are significantly higher than that for case B (0.565).
We also find that
ULIRGs have a tendency to exhibit higher Br$\beta$/Br$\alpha$
line ratios than those observed in Galactic \ion{H}{2} regions.
If dust extinction affects the flux of the lines,
the Br$\beta$/Br$\alpha$ line ratio will decrease below 0.565
because the Br$\beta$ line has a shorter wavelength
and so is more attenuated than the Br$\alpha$ line.
Thus, we cannot explain the high Br$\beta$/Br$\alpha$ line ratio
with a combination of case B theory and dust extinction.

We investigated the cause of this anomaly and
obtained the following results:

\begin{enumerate}
\item
We explored the possibility
of contamination of the Brackett lines by other lines.
We identified one candidate, the H$_2$ (1,0) O(2)
line, with a wavelength of 2.627~$\mu$m,
that is close to the wavelength of the Br$\beta$ line (2.626~$\mu$m).
We estimated the flux of the H$_2$ (1,0) O(2) line
from that of another molecular hydrogen line, H$_2$ (1,0) O(3)
at 2.802~$\mu$m, assuming that the line ratio
corresponds to that of a $2000$~K shock model \citep{Black1987fei}.
The expected flux of the H$_2$ (1,0) O(2) line
is 5--12\% of that of the observed Br$\beta$ line.
Consequently, the Br$\beta$/Br$\alpha$ line ratio is still
more than 3$\sigma$ higher than
that for case~B
in IRAS~10565$+$2448 and Mrk~273
even after we subtract the
flux of the H$_2$ (1,0) O(2) line
from that of the Br$\beta$ line.
Thus, we conclude that contamination does not provide
a complete explanation of the high Br$\beta$/Br$\alpha$ line ratio.
\item
For the case in which the Brackett lines are optically thin,
we cannot explain the high Br$\beta$/Br$\alpha$ line ratio
with any of the three possible excitation mechanisms:
recombination, collisional excitation, or resonant excitation.
\item
We find that we can explain the deviation of the Br$\beta$/Br$\alpha$ line ratio
from that of case B if the Br$\alpha$ line becomes optically thick
while the Br$\beta$ line is still optically thin.
\item
We simulated \ion{H}{2} regions, each ionized by a single star,
with the Cloudy code
and found that the high Br$\beta$/Br$\alpha$ line ratio
can be explained when the Br$\alpha$ line becomes optically thick.
To achieve a column density large enough
to make the Br$\alpha$ line optically thick within a single \ion{H}{2} region,
the gas density must be as high as $n\sim10^8$~cm$^{-3}$.
\item
We investigated the ratios of optical \ion{H}{1} lines
in the galaxies in our sample that show high Br$\beta$/Br$\alpha$ line ratios.
We found that the fluxes of the optical lines are highly affected
by dust extinction,
and it is difficult to tell whether the line ratio contradicts
case~B theory.
We conclude that the deviation of the \ion{H}{1} line ratios from those of case B
can be seen clearly only in the infrared \ion{H}{1} lines
because the optical lines are strongly affected by dust extinction.
\item
We investigated the consistency of our high-density model
with other infrared \ion{H}{1} line observations.
We compared the \ion{H}{1} line ratios other than Br$\beta$/Br$\alpha$
with those predicted by the high-density model
for the three galaxies with high Br$\beta$/Br$\alpha$ line ratios.
We conclude that our model is consistent with
previous observations of the Pa$\alpha$/Br$\gamma$ line ratio.
\item
We consider two extreme cases for the line-emitting regions:
(1) an ensemble of \ion{H}{2} regions, each ionized by a single star
and (2) a giant \ion{H}{2} region where all the ionizing stars are
concentrated at the center.
For both the cases,
we conclude that gas densities as high as $n\sim10^8$~cm$^{-3}$
are required to achieve a column density of neutral hydrogen large enough
to make the Br$\alpha$ line optically thick.
We propose this high-density scenario as the most plausible cause
of the high Br$\beta$/Br$\alpha$ line ratio.
The required density may be increased by up to an order of magnitude
if dust grains are taken into account.
\item
Our model requires high-density \ion{H}{2} regions
with $n=10^8$~cm$^{-3}$.
This affects the high-$\mathcal{N}$ transitions of \ion{H}{1} lines,
which fall in the radio-frequency range.
We predict that radio recombination lines with $\mathcal{N}<50$
are thermalized in galaxies with high Br$\beta$/Br$\alpha$ line ratios.
\end{enumerate}

\acknowledgments

We thank the anonymous referee
for reading our paper carefully and sending many useful suggestions for improvement.
This study is based on the observations made with
\textit{AKARI}, a JAXA project, with the participation of ESA.
We also thank the Lick Observatory staff
for their assistance.
This research made use of
the NASA/IPAC Extragalactic Database, which is operated
by the Jet Propulsion Laboratory, California Institute of
Technology, under contract with the National Aeronautics and
Space Administration.
Data analysis was in part carried out
on the Multi-wavelength Data Analysis System 
operated by the Astronomy Data Center (ADC),
National Astronomical Observatory of Japan.
This work is supported by JSPS KAKENHI Grant Number 26247030.
K.Y.~is supported through the Leading Graduates
Schools Program, Advanced Leading Graduate Course for Photon Science, by the Ministry of
Education, Culture, Sports, Science and Technology of Japan.
S.B.~is supported by JSPS KAKENHI Grant Number JP19J00892.

\facilities{\textit{AKARI}(IRC), Nickel: 1.0~m}

\software{
Cloudy \citep[ver.~10.00;][]{Ferland1998c9n},
IRC Spectroscopy Toolkit \citep{Ohyama2007nia,Baba2016r},
IPython \citep{IPython},
Jupyter Notebook \citep{JupyterNotebook},
NumPy \citep{NumPy},
SciPy \citep{SciPy},
Pandas \citep{Pandas},
Matplotlib \citep{Matplotlib},
Astropy \citep{Astropy,Astropy_v2},
Lmfit \citep{lmfit}
}

\appendix

\begin{deluxetable}{lcc}
\tablecaption{Observation Log for NULIZ Targets \label{tab:id}}
\tablewidth{0pt}
\tablehead{
\colhead{Object Name}&\colhead{Observation ID}&\colhead{Observation Date}
}
\startdata
IRAS 09022$-$3615&3051018.1&2007 May 26\\
IRAS 10565$+$2448&3051019.1&2007 May 28
\enddata
\end{deluxetable}

\begin{deluxetable*}{ccccccccc}
\tablecaption{Basic Information for NULIZ Targets \label{tab:info}}
\tablewidth{0pt}
\tabletypesize{}
\tablehead{
\colhead{Object Name}&\colhead{$z$\tablenotemark{a}}&\colhead{$D_\mathrm{L}$\tablenotemark{b}}&\colhead{$F_{25}$\tablenotemark{c}}&\colhead{$F_{60}$\tablenotemark{c}}&\colhead{$F_{100}$\tablenotemark{c}}&\colhead{$L_\mathrm{IR}$\tablenotemark{d}}&\colhead{Optical\tablenotemark{e}}&\colhead{Ref.\tablenotemark{f}}\\
\colhead{}&\colhead{}&\colhead{(Mpc)}&\colhead{(Jy)}&\colhead{(Jy)}&\colhead{(Jy)}&\colhead{($10^{12}L_\odot$)}&\colhead{class}&\colhead{}
}
\startdata
IRAS 09022$-$3615&0.060&266&1.20&11.6&11.1&1.64&\ion{H}{2}&1\\
IRAS 10565$+$2448&0.043&190&1.27&12.1&15.0&1.07&\ion{H}{2}&2
\enddata
\tablenotetext{a}{Redshift.}
\tablenotetext{b}{\mbox{}Luminosity distance calculated from $z$ using our adopted cosmology.}
\tablenotetext{c}{\textit{IRAS} fluxes at 25~$\mu$m ($F_{25}$), 60~$\mu$m ($F_{60}$),
and 100~$\mu$m ($F_{100}$).}
\tablenotetext{d}{Total infrared (3--1100~$\mu$m) luminosity calculated with
$L_\mathrm{IR}=4\pi D_\mathrm{L}^2(\xi_1\nu F_{25}+\xi_2\nu F_{60}+\xi_3\nu F_{100})$, where
$(\xi_1,\ \xi_2,\ \xi_3)=(2.403,\ -0.2454,\ 1.6381)$ \citep{Dale2002ise}.}
\tablenotetext{e}{Optical classification of galaxies.}
\tablenotetext{f}
{References for redshifts:
(1) \cite{Strauss1992rsi};
(2) \cite{Downes1993mgm}.}
\end{deluxetable*}

\begin{deluxetable*}{cccc}
\tablecaption{Flux Ratios of the H$\alpha$ and H$\beta$ Lines \label{tab:hbha}}
\tablewidth{0pt}
\tabletypesize{}
\tablehead{
\colhead{Object Name}&\colhead{$F_{\mathrm{H}\beta}/F_{\mathrm{H}\alpha}$}&\colhead{$A_V^\mathrm{opt}$\tablenotemark{a}}&\colhead{Reference\tablenotemark{b}}\\
\colhead{}&\colhead{}&\colhead{(mag)}&\colhead{}
}
\startdata
IRAS 09022$-$3615&0.18$\pm$0.009&$1.90\pm0.14$&1\\
IRAS 10565$+$2448&0.06$\pm$0.003&$4.83\pm0.14$&2\\
\enddata
\tablenotetext{a}{Visual extinction derived from H$\alpha$/H$\beta$ line ratio.}
\tablenotetext{b}
{References for optical line ratio:
(1) \cite{Lee2011osc};
(2) \cite{Veilleux1995osl}.}
\end{deluxetable*}

\section{NULIZ Targets}
\label{sec:nuliz}

We summarize here the properties of
the two \textit{AKARI} targets added from the NULIZ program.
The observation log and basic information are
listed in Tables~\ref{tab:id} and \ref{tab:info}.
The H$\beta$/H$\alpha$ line ratio
used in \S\ref{sec:cwh}
is summarized in Table~\ref{tab:hbha}.

\section{Line Ratio in the Galactic \ion{H}{2} Regions}
\label{sec:gal}

Our results for the high Br$\beta$/Br$\alpha$ line ratio
indicate that conditions in the \ion{H}{2} regions
in the ULIRGs that exhibit the anomalies are different from
those of Galactic \ion{H}{2} regions,
where the case~B theory explains the line ratios well.
To investigate the difference,
we here discuss the applicability of the case~B line ratio for the Br$\alpha$ and Br$\beta$ lines
in Galactic \ion{H}{2} regions.

Because of the difficulty of observing the wavelength range where
the Br$\alpha$ and Br$\beta$ lines lie due to the atmospheric absorption,
previous observations have been
limited to those conducted in space,
namely, observations with the \textit{ISO} and \textit{AKARI} satellites.
Accordingly, we investigated observations of Galactic \ion{H}{2} regions
obtained with these two satellites.

\subsection{\textit{ISO} Observations}

\cite{Lutz1996sog} obtained a 2.4--45 $\mu$m
spectrum of the Galactic center
with the Short Wavelength Spectrometer on board \textit{ISO}.
The $14''\times21''$ aperture was centered on Sgr~A$^*$
to cover \ion{H}{2} regions in the Galactic center region.
They detected \ion{H}{1} lines, including the Br$\alpha$ and Br$\beta$ lines,
in the wavelength range 2.5--9.0~$\mu$m.
The Br$\beta$/Br$\alpha$ line ratio at the Galactic center
was reported to be $\sim0.25$,
which is consistent with case~B
plus dust extinction \citep{Lutz1999iog}.
\cite{Lutz1999iog} also discussed the applicability
of case~B in the Galactic center
using the Br$\alpha$ line,
the Pf$\alpha$ line ($\mathcal{N}=6\rightarrow5$; 7.46~$\mu$m),
and a blend of the Hu$\beta$ line ($\mathcal{N}=8\rightarrow6$; 7.50~$\mu$m)
with the $\mathcal{N}=11\rightarrow7$ transition (7.51~$\mu$m).
They concluded that the flux ratios of these lines
were consistent with the case~B line ratios
and that the population of the respective upper levels
followed case~B.
Thus, the \textit{ISO} results indicate that
the case~B line ratio is applicable to \ion{H}{2} regions near the Galactic center.

\subsection{\textit{AKARI} Observations}

\begin{figure}
\centering
\includegraphics[width=\columnwidth]{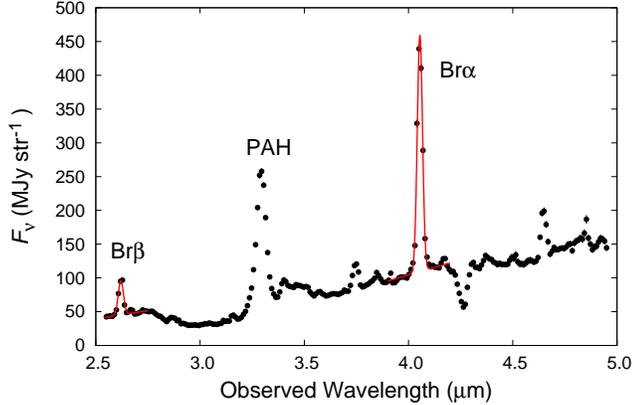}
\caption{Typical example of a cataloged spectrum
of a Galactic \ion{H}{2} region by \cite{Mori2014osn}.
The spectrum was obtained from the position
$-$04 of W31a (ID: 5200165.1) using the
``Nh'' slit.
The best-fit Gaussian profile for the Br$\alpha$ and Br$\beta$ lines
are shown as red curves.
\label{fig:hex}}
\end{figure}

Using \textit{AKARI}
near-infrared spectroscopy,
\cite{Mori2014osn} conducted a systematic observation
of 36 Galactic \ion{H}{2} regions
and provided a catalog of 2.5--5.0$\mu$m spectra
of such objects.\footnote{The catalog is publicly available
at URL: \url{http://www.ir.isas.jaxa.jp/AKARI/Archive/Catalogues/IRC_GALHII_spec/}}
A typical example from the cataloged spectra is shown in Figure~\ref{fig:hex}.

We determined the Br$\alpha$ and Br$\beta$ line fluxes
for those Galactic \ion{H}{2} regions using the 232 cataloged spectra.
We fitted the Br$\alpha$ and Br$\beta$ lines separately, with
a Gaussian profile and a linear continuum for each spectrum.
Following \cite{Mori2014osn},
we fixed the FWHM of the Gaussian profile
at 0.031~$\mu$m for spectra taken with the ``Ns'' slit
and 0.025~$\mu$m for those taken with ``Nh'' slit
in order to match the spectral resolution of the slits.
The central wavelengths of the lines were also fixed
at 4.05~$\mu$m for the Br$\alpha$ line
and at 2.63~$\mu$m for the Br$\beta$ line.
The range of wavelengths used for the fitting
was $\pm0.15$~$\mu$m around the central
wavelength of each line.
We then determined the line flux by integrating the best-fit Gaussian profile.
A typical example of the Gaussian fitting is shown in Figure~\ref{fig:hex}.

\begin{figure}
\centering
\includegraphics[width=\columnwidth]{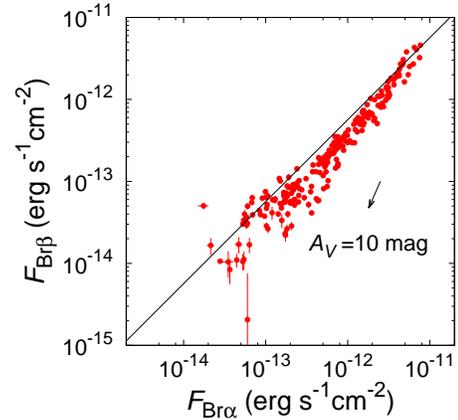}
\caption{The fluxes of the Br$\alpha$ and Br$\beta$ lines for 
Galactic \ion{H}{2} regions obtained from
the spectral catalog of \cite{Mori2014osn}.
The solid line shows the theoretical line ratio for case B:
$F_{\mathrm{Br}\beta}/F_{\mathrm{Br}\alpha}=0.565$.
The extinction vector corresponding to $A_V=10$ mag is shown as the black arrow.
\label{fig:hiiregion}}
\end{figure}

We show the fluxes of
the Br$\alpha$ and Br$\beta$ lines obtained for the Galactic \ion{H}{2} regions
in Figure~\ref{fig:hiiregion}.
The results show that almost all the Br$\beta$/Br$\alpha$
line ratios are lower than that for case~B,
except for a few spectra.
This indicates that the Br$\beta$/Br$\alpha$ line
ratios of Galactic \ion{H}{2} regions can be explained by case~B condition
plus dust extinction, which is typically $A_V\sim10$~mag.

Combining the results of the \textit{ISO} and \textit{AKARI} observations,
we conclude that the case~B condition is valid for Galactic \ion{H}{2} regions.
Thus, the anomalous Br$\beta$/Br$\alpha$
line ratios found in some ULIRGs indicate that conditions in the \ion{H}{2} regions
in those ULIRGs differ from the case~B conditions.

\section{Detailed Discussions on Resonant Processes}

This appendix provides detailed discussions
about the effect of resonant excitations
on the Br$\beta$/Br$\alpha$ line ratio.
First, we discuss
the $\mathcal{N}=1\rightarrow6$ resonance
referred in \S\ref{sec:thin}.
Then, we consider
the $\mathcal{N}=1\rightarrow4$ resonance
mentioned in \S\ref{sec:pcp}.

\subsection{Resonant Excitation to the $\mathcal{N}=6$ State}
\label{sec:ap6}

With resonant excitation,
the hydrogen $\mathcal{N}=6$ state is enhanced
if a strong line with a wavelength
comparable to that of the transition
$\mathcal{N}=6\rightarrow1$ (937.8~\AA) exists.
We denote this (unknown) line by X$_{6}$
and discuss its effect on the Br$\beta$/Br$\alpha$ line ratio
in the optically thin case described in \S\ref{sec:thin}.

First, we assume a case in which hydrogen is
excited solely by this resonant excitation.
Let $x_6$ be the number of resonance excitations
per unit volume and unit time (cm$^{-3}$~s$^{-1}$).
Then, the number of Br$\beta$ transitions ($\mathcal{N}=6\rightarrow4$)
caused by the resonant excitation can be 
written as $x_6A_{6,4}/A_6\sim0.22x_6$,
and the number of Br$\alpha$ transitions ($\mathcal{N}=5\rightarrow4$)
is given by $x_6C_{6,5}C_{5,4}=0.11x_6$.
Here, the $A$ and $C$ symbols are the Einstein A coefficient and cascade matrix
introduced in \S\ref{sec:reco}, respectively.
The Br$\beta$/Br$\alpha$ line ratio is thus found to be
$F_{\mathrm{Br}\beta}/F_{\mathrm{Br}\alpha}
=0.22x_6h\nu_{\mathrm{Br}\beta}/0.11x_6h\nu_{\mathrm{Br}\alpha}\sim3.2$.
Therefore, it is possible for resonant excitation to make the line ratio
consistent with our observations,
given the existence of an appropriate line transition X$_{6}$.

Next we discuss how strong the X$_6$ line
must be in order explain the observed anomaly.
Taking the resonant excitation rate $x_6$ into account
in Equation~(\ref{eq:nstate}),
we now write the level populations for $\mathcal{N}=5$ and 6 as
\begin{eqnarray}
&&n_6'A_6=
n_\mathrm{p}n_\mathrm{e}\sum^\infty_{\mathcal{N}'=6}
\alpha_{\mathcal{N}'}C_{\mathcal{N}',6}+x_6,\nonumber\\
&\therefore&\quad
n_6'=n_\mathrm{p}n_\mathrm{e}\frac{\sum^\infty_{\mathcal{N}'=6}
\alpha_{\mathcal{N}'}C_{\mathcal{N}',6}}{A_6}+\frac{x_6}{A_6},\label{eq:n6x}
\end{eqnarray}
and
\begin{eqnarray}
&&n_5'A_5=
n_\mathrm{p}n_\mathrm{e}\sum^\infty_{\mathcal{N}'=6}
\alpha_{\mathcal{N}'}C_{\mathcal{N}',5}+C_{6,5}x_6,\nonumber\\
&\therefore&\quad
n_5'=n_\mathrm{p}n_\mathrm{e}\frac{\sum^\infty_{\mathcal{N}'=5}
\alpha_{\mathcal{N}'}C_{\mathcal{N}',5}}{A_5}+\frac{C_{6,5}x_6}{A_5}.\label{eq:n5x}
\end{eqnarray}
In order to explain the observed anomaly, in which
$F_{\mathrm{Br}\beta}/F_{\mathrm{Br}\alpha}\sim1$,
we require the ratio of the level populations
to be $n_6'/n_5'\sim2.27$.
Using Equations~(\ref{eq:n6x}) and (\ref{eq:n5x}),
we find
\begin{eqnarray}
&&n_\mathrm{p}n_\mathrm{e}\frac{\sum^\infty_{\mathcal{N}'=6}
\alpha_{\mathcal{N}'}C_{\mathcal{N}',6}}{A_6}+\frac{x_6}{A_6} \nonumber\\
&&\quad=2.27
\left(n_\mathrm{p}n_\mathrm{e}\frac{\sum^\infty_{\mathcal{N}'=5}
\alpha_{\mathcal{N}'}C_{\mathcal{N}',5}}{A_5}+\frac{C_{6,5}x_6}{A_5}\right),\nonumber\\
&\therefore&\quad
x_6=2.29\times10^{-14}\left(\frac{n}{\mathrm{cm}^{-3}}\right)^2\ \mathrm{cm}^{-3}\mathrm{s}^{-1},\label{eq:xlim}
\end{eqnarray}
where we have approximated $n_\mathrm{e} n_\mathrm{p}\sim n^2$.
Assuming that all photons emitted in the X$_6$ line
are absorbed by the hydrogen atoms,
we can obtain that $x_6$ is equal to the emission rate of the X$_6$ line.
We write this rate as $n\xi_\mathrm{X}f_{\mathrm{X}_6}A_{\mathrm{X}_6}$,
where $\xi_\mathrm{X}$ is the abundance of the atoms emitting the X$_{6}$ line, relative to hydrogen;
$f_{\mathrm{X}_6}$ is the fraction of excited atoms that can radiate the X$_{6}$ line,
relative to those in other states; and $A_{\mathrm{X}_6}$ is the Einstein A coefficient for the X$_{6}$ line.
From Equation~(\ref{eq:xlim}), we thus obtain
\begin{equation}
A_{\mathrm{X}_6}=2.29\times10^{-14}f_{\mathrm{X}_6}^{-1}\xi_\mathrm{X}^{-1}
\left(\frac{n}{\mathrm{cm}^{-3}}\right)\ \mathrm{s}^{-1}.
\end{equation}
It is difficult to estimate the fraction $f_{\mathrm{X}_6}$, so we here
assume the most extreme case that it is of order unity
in order to take weak lines into consideration.
At a gas density $n=10^3$~cm$^{-3}$
and assuming that the atoms emitting the X$_{6}$ line have
an abundance similar to those of the metals,
$\xi_\mathrm{X}\sim10^{-4}$,
we find $A_{\mathrm{X}_6}$ to be
$\sim10^{-7}$~s$^{-1}$,
which is close to the values for forbidden lines.
This indicates that if a line exists with a wavelength
$\sim$937.8~\AA\ and a transition probability
comparable to those of forbidden lines,
then the resonant process
would be able to make the Br$\beta$/Br$\alpha$ line
ratio anomalously high.

\subsection{Resonant Excitation to the $\mathcal{N}=4$ State}
\label{sec:ap4}

Herein, we discuss
the effect of the $\mathcal{N}=1\rightarrow4$ resonant excitation
on the Br$\beta$/Br$\alpha$ line ratio in the optically thick case
described in \S\ref{sec:pcp}.
We denote the possible resonant line by X$_4$
and write the rate of resonant excitation
by this line as $x_4$ (cm$^{-3}$~s$^{-1}$).
From Equation~(\ref{eq:nstate}),
the $\mathcal{N}=4$ state is populated by
the recombination process at the rate
$n_\mathrm{e}n_\mathrm{p}
\sum_{\mathcal{N}'=4}^\infty\alpha_{\mathcal{N}'}
C_{\mathcal{N}',4}\sim6.2\times10^{2}$~cm$^{-3}$~s$^{-1}$ at $n=10^8$~cm$^{-3}$.
This indicates that if $x_4$ is larger than
$\sim10^3$~cm$^{-3}$~s$^{-1}$,
the resonant process can significantly populate
the $\mathcal{N}=4$ state
when $n=10^8$~cm$^{-3}$.

In the same way as for the hypothetical
$\mathcal{N}=1\rightarrow6$ resonance
we discussed in \S\ref{sec:ap6},
we can estimate the transition probability $A_{\mathrm{X}_4}$
of the X$_4$ line that is
required to make the resonant process
dominant for the $\mathcal{N}=4$ population.
We assume that all photons emitted by the X$_4$ line
are absorbed by hydrogen
such that $x_4$ is equal to the rate of
emission of the X$_4$ line,
$n\xi_\mathrm{X}f_{\mathrm{X}_4}A_{\mathrm{X}_4}$,
where $\xi_\mathrm{X}$ is the abundance of
the atoms emitting the X$_4$ line relative to hydrogen,
and $f_{\mathrm{X}_4}$ is the fraction of excited atoms which can radiate the X$_{4}$ line
relative to those in other states.
We assume $f_{\mathrm{X}_4}$
to be of the order of unity as the most extreme case.
At a gas density of $n=10^8$~cm$^{-3}$
and assuming that the atoms emitting the X$_{4}$ line have
an abundance similar to those of the metals,
i.e., $\xi_\mathrm{X}\sim10^{-4}$,
we estimate $A_{\mathrm{X}_4}$ to be
$\sim10^{-1}$~s$^{-1}$.
Thus, if a line with a wavelength of
$\sim$972.5~\AA\ and a transition probability
of $\sim10^{-1}$~s$^{-1}$
exists, we should take the resonant process into consideration 
in determining the population of the $\mathcal{N}=4$ state.

\bibliography{yano_anom}{}
\bibliographystyle{aasjournal}

\end{document}